\documentclass[12pt]{article}
\usepackage{subfigure}
\usepackage{amsmath}
\usepackage{amsfonts}
\usepackage{amssymb}
\usepackage{color}
\usepackage{calc}
\usepackage[latin1]{inputenc}
\usepackage[english]{babel}
\usepackage[T1]{fontenc}
\usepackage{graphicx}
\usepackage{verbatim}
\usepackage{subfig}
\usepackage{enumerate}
\usepackage[numbers]{natbib}

\def\Bmp#1{ \begin{minipage}{#1} }
\def\Bmpc#1{ \begin{minipage}[c]{#1} }
\def\Bmpt#1{ \begin{minipage}[t]{#1} }
\def\Bmpb#1{ \begin{minipage}[b]{#1} }
\def\Emp{ \end{minipage} }
\newcommand{\revt}[1]{{\color{black}#1}}

\def\I{{\mathcal{I}}}
\def\J{{\mathcal{J}}}

\def\L{{\mathcal{L}}}
\def\X{{\mathcal{X}}}
\def\O{{\mathcal{O}}}

\def\R{{\mathcal{R}}}

\def\tf0{\tilde{\varphi}_{0}}

\def\RR{{\mathbb{R}}}

\def\e{{\bf e}}
\def\n{{\bf n}}

\def\x{{\bf x}}

\def\0{{\bf 0}}

\def\bnabla{\boldsymbol{\nabla}}
\def\bomega{\boldsymbol{\omega}}

\def\Dpartial#1#2{ {\frac{\partial #1}{\partial #2} }}

\newcommand{\argmin}{\operatorname{argmin}}

\newtheorem{problem}{Problem}

\usepackage{color}

\newcommand{\ie}{{\em i.\thinspace{}e. }}
\newcommand{\eg}{{\em e.\thinspace{}g. }}

\def\vect3#1#2#3{\displaystyle \left(#1, #2, #3\right)^T}

\def\vec#1{{\bf #1}}

\newcommand{\pl}{\partial }

\begin{document}

\title{Optimal Reconstruction of Inviscid Vortices}
\author{Ionut Danaila$^1$ and Bartosz Protas$^2$
\\ \\
$^1$Laboratoire de math\'ematiques Rapha\"el Salem \\
Université de Rouen \\
Technop\^ole du Madrillet \\
76801 Saint-\'Etienne-du-Rouvray, FRANCE \\
\\ \\
$^2$Department of Mathematics \& Statistics, \\
McMaster University  \\
Hamilton, Ontario L8S4K1, CANADA }

\date{\today}
\maketitle

\begin{abstract}
  We address the question of constructing simple inviscid vortex
  models which optimally approximate realistic flows as solutions of
  an inverse problem. Assuming the model to be incompressible,
  inviscid and stationary in the frame of reference moving with the
  vortex, the "structure" of the vortex is uniquely characterized by
  the functional relation between the streamfunction and vorticity.
  It is demonstrated how the inverse problem of reconstructing this
  functional relation from data can be framed as an optimization
  problem which can be efficiently solved using variational
  techniques.  In contrast to earlier studies, the vorticity function
  defining the streamfunction-vorticity relation is reconstructed in
  the continuous setting subject to a minimum number of assumptions.
  To focus attention, we consider flows in 3D axisymmetric geometry
  with vortex rings. To validate our approach, a test case involving
  Hill's vortex is presented in which a very good reconstruction is
  obtained.  In the second example we construct an optimal inviscid
  vortex model for a realistic flow in which a more accurate vorticity
  function is obtained than produced through an empirical fit.  When
  compared to available theoretical vortex-ring models, our approach
  has the advantage of offering a good representation of both the
  vortex structure and its integral characteristics.  
\end{abstract}

\begin{flushleft}
\end{flushleft}

\tableofcontents

\section{Introduction}
\label{sec:intro}

\begin{figure}[!htpb]
\begin{center}
\includegraphics[width=1.0\textwidth]{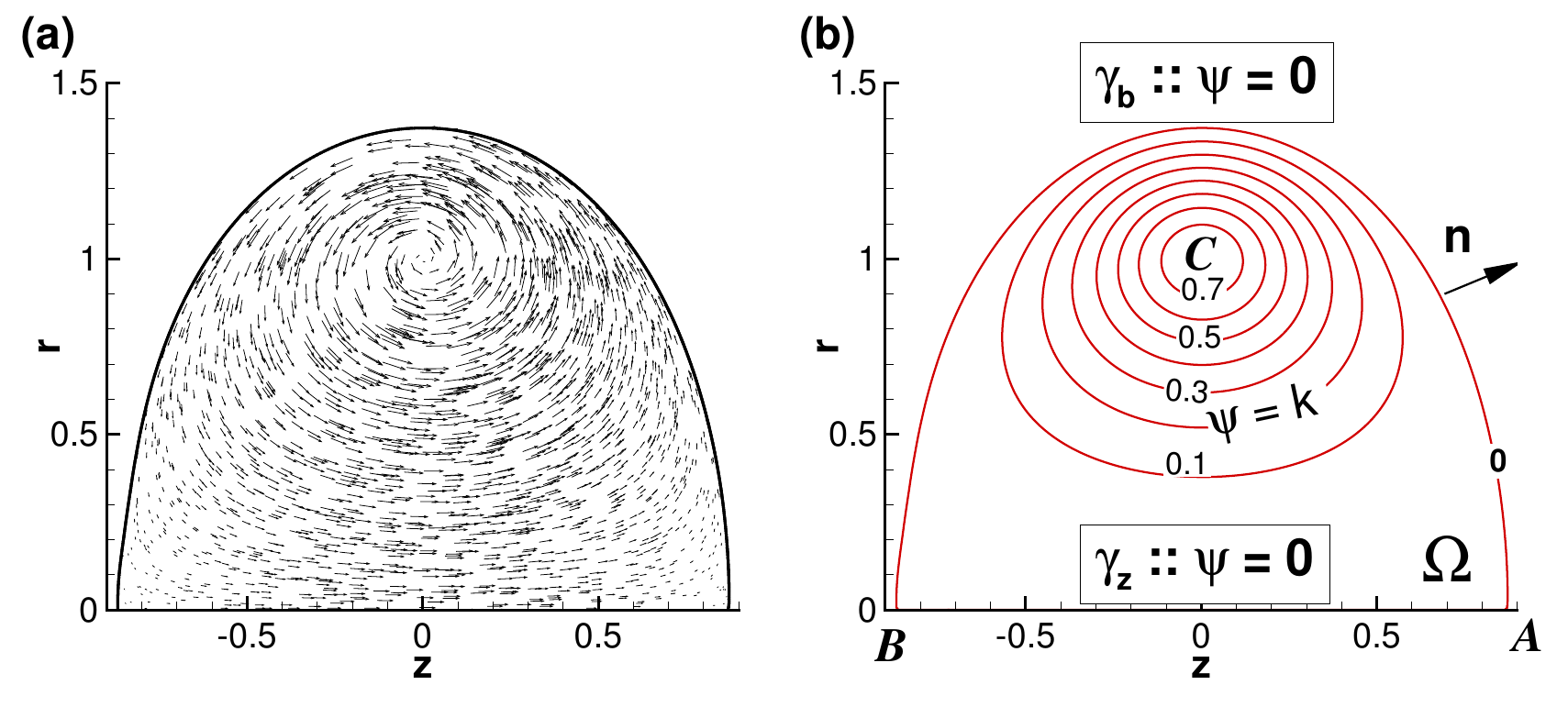}
\vspace*{-0.75cm}
\end{center}
\caption{Direct numerical simulation of the incompressible
  Navier-Stokes equations representing a physical vortex ring with
  axial symmetry \cite{dan-2008}. Velocity vectors (a) and
  corresponding streamlines (b) in the frame of reference travelling
  with the vortex ring.}
\label{fig-intro-dns}
\end{figure}

{In this investigation we study the problem of constructing
  simple inviscid models for flows of realistic incompressible fluids.
  More specifically, we are interested in situations where such flows
  can be approximately represented by localized vortices which are
  steady in a suitable frame of reference. As an example of this type
  of flow phenomena, vortex rings are ubiquitous in many important
  applications ranging from biological propulsion
  \cite{mohseni-2006,dabiri-2009-AR} to the fuel injection in internal
  combustion engines \cite{begg-2009-ER,kaplanski-2010-EJMB}. We will
  thus focus on constructing steady inviscid incompressible flows
  which in some mathematically precise sense provide an optimal
  representation of the original flow field. Such Euler flows are
  described by equations of the type $\L \psi = F(\psi)$, where $\L$
  is a second-order self-adjoint elliptic operator specific to the
  particular flow configuration and $\psi$ represents the
  streamfunction in two dimensions (2D) and the Stokes streamfunction
  in three dimensions (3D). The nonlinear source function $F(\psi)$
  encodes information about the structure of the inviscid vortex.
  Therefore, the question of identifying an inviscid vortex best
  matching a given velocity field leads to an {\em inverse problem}
  for the reconstruction of the source function $F(\psi)$. While there
  have been numerous attempts to model realistic flows in terms of
  localized vortices, especially vortex rings
  \cite{kaplanski-1999-IJFMR,kaplanski-2005-PF,fukumoto-2010}, the
  idea of framing this as an inverse problem has received rather
  little attention with earlier approaches relying on the
  representation of the unknown source function in terms of a small
  number of parameters \cite{dan-2012-JNM}. In the present study we
  propose and validate a fundamentally different approach which will
  allow us to reconstruct the source function $F(\psi)$ in a very
  general form as a continuous function subject to minimal
  assumptions. This approach is an adaptation of the method for an
  optimal reconstruction of constitutive relations developed in
  \cite{bvp10,bp11a} which was recently also used to study a number of
  other problems in fluid mechanics \cite{pnm14,pno14a}. Given that
  inverse problems for partial differential equations (PDEs) are often
  ill-posed \cite{i98}, another objective of the present study is to
  assess to what extent such reconstruction is actually possible for
  selected problems and identify its limitations. This will also
  provide insights about physical aspects of the problem which are
  captured by the reconstruction approach.

  To fix attention, but without the loss of generality, hereafter we
  will focus on axisymmetric flows in 3D geometry with vortex rings
  (Figure \ref{fig-intro-dns}). A very similar approach can be
  developed for 2D flows.}  For convenience, {in the following we use
  the} cylindrical coordinates $(z,r,\theta)$ with $z$ the
longitudinal (propagation) direction of the flow. We denote {the
  velocity field by} $\vec{v}=\vect3{v_z}{v_r}{0}$ and by
$\bomega=\bnabla \times \vec{v} = \vect3{0}{0}{\omega_\theta}$ the
corresponding vorticity. {Thus, given our assumption that the flow is
  axisymmetric,} the only nonzero vorticity component is the azimuthal
one, denoted by $\omega := \omega_\theta$.

A vortex ring is then defined as the axisymmetric region $\Omega_b$ of
$\RR^3$ such that $\omega \neq 0$ in $\Omega_b$ and $\omega = 0$
elsewhere {(see Figure \ref{fig-intro-dns})}. The domain $\Omega_b$,
also called {\it vortex bubble}, is delimited by {the streamline
  corresponding to} $\psi=0$, where $\psi(z,r)$ is the Stokes
streamfunction in the frame of reference moving with the vortex ring.

Classical vortex ring models are stationary solutions of Euler
equations. The key feature of such models is that the vorticity
transport equation reduces to (\eg \cite{batchelor-1988,saffman-1992})
\begin{equation} 
\frac{\omega}{r} = \left\{
\begin{tabular}{ll}
$f(\psi)$ &  \mbox{in} \,\, $\Omega_b$,\\
0       & \mbox{elsewhere},
\end{tabular}
\right.
\label{eq-om-fpsi}
\end{equation}
with $f : \RR \rightarrow \RR$ called the {\it vorticity function}
{(it is related to the source function introduced above as
  $F(\psi) = r f(\psi)$)}. In other words, $\omega$ propagates with a
time-invariant profile {$f(\psi)$} along streamlines in the frame of
reference moving with the vortex.  {As described in detail in the next
  section,} the associated mathematical problem consists in solving an
elliptic PDE for $\psi$ with a right-hand-side term depending on the
solution itself. The main difficulty in solving this PDE comes from
the fact that the boundary $\partial\Omega_b$ {of the vortex bubble}
is not known {in advance}, which makes it a {\em free boundary
  problem}.

The only known analytical solution of this problem considers
$f(\psi)=const$ {within $\Omega_b$ which has the shape of a sphere},
{and} is known as Hill's spherical vortex \cite{hill-1894} (see
also \cite{batchelor-1988,saffman-1992}).  The mathematical theory of
inviscid axisymmetric vortex rings was developed in the '70s and in
the early '80s \cite{fraenkel-1974-AR} around Hill's vortex, by
{considering the following particular form of the vorticity function}
\begin{equation}
f(\psi) = const, \quad \forall \psi > k, \qquad \mbox{and} \qquad f(\psi) = 0, \quad \forall \psi \leq k,
 \label{eq-gen-frz}
\end{equation}
with $k>0$ defining the {\it vortex-ring core} as $\Omega_c=\left\{
  ({z, r, \theta}) \in \RR^{3}, \psi(z,r) > k \right\} \subset
\Omega_b$ (see Figure \ref{fig-intro-dns}b).  Existence and uniqueness
results for the inviscid vortex ring problem are presented in
\cite{fraenkel-1974-AR,ni-1980,amick-1988} for the general case and in
\cite{amick-1986,norbury-1972-proc} for vortex rings bifurcating from
Hill's vortex. Numerical solutions of the vortex-ring problem, using
\eqref{eq-gen-frz} as {the} vorticity function, were {obtained by
  Norbury \cite{norbury-1973-JFM} and Fraenkel
  \cite{fraenkel-1972-JFM}, and are hereafter referred to as NF
  vortices.}  These models were also extended to allow for
{vortex rings with swirl} (i.e., with nonzero azimuthal
velocity); analytical closed-form solutions for Hill's spherical
vortex {with swirl} were obtained in \cite{moffatt69} and
numerical solutions {using} vorticity functions generalizing
\eqref{eq-gen-frz} for swirling flows were presented in
\cite{elcrat-2008}.

From the practical point of view, vortex-ring models are useful as
{inviscid} approximations to actual vortex structures observed in
experiments or generated by the Direct Numerical Simulation (DNS) of
{the} Navier-Stokes equations.  {For the purposes of fitting such
  models to DNS data}, the NF inviscid vortex-ring model
\cite{fraenkel-1972-JFM,norbury-1973-JFM} was {widely adopted} and
proved very useful in estimating integral quantities and global
properties of actual vortex rings
\cite{mohseni-1998-PF,shusser-2000-PF,linden-2001}. This is quite
remarkable, since the vorticity function \eqref{eq-gen-frz} gives a
linear vorticity distribution in the vortex core, \ie proportional to
the distance from the axis of symmetry, which is in fact quite
different from the Gaussian vorticity distribution typically observed
in experiments (\eg \cite{weigand-1997-ExpF,cater-2004}).  The main
feature of the {inviscid} vortex-ring models is that the
vorticity function $f(\psi)$ is prescribed by
{\eqref{eq-om-fpsi}} as a hypothesis of the model. While
experimental studies \cite{sullivan-1973,akhmetov-2009} reported some
{scatter in the plots of $\omega/r$ versus $\psi$, this data was
  rather well fitted by an empirical formula for the vorticity
  function in the exponential form $f(\psi) = a \exp(b\psi)$ with $a$
  and $b$ representing two constants adjusted during the fitting
  procedure \cite{akhmetov-2009}.} This supports the idea that steady
inviscid models could be used as good approximations of unsteady
viscous vortex rings {arising in real flows} if the vorticity function
$f(\psi)$ is accurately determined.

{In the present contribution we formulate the problem of
  identifying an optimal vorticity function $f(\psi)$ as an inverse
  problem. It will be solved using a variational optimization approach
  in which optimality of the reconstruction implies that the obtained
  inviscid vortex-ring model best matches, in a suitably defined
  sense, the available measurements. A natural question in the
  formulation of such an inverse problem is how much measurement data
  is required to ensure a reliable reconstruction The solution method
  we have developed can assimilate measurements available in 3D
  regions or on 2D surfaces with in principle arbitrary shapes. While
  modern experimental techniques such as particle-image velocimetry
  (PIV) and advanced DNS can provide snapshots of the velocity field
  in a large part of the flow domain, for benchmarking purposes in our
  computational examples we will consider an arguably harder problem
  where only incomplete measurements are available. More specifically,
  we will assume the reconstructions to be based on measurements of
  the tangential velocity component on $\partial\Omega_b$, the
  boundary of the vortex bubble. A practical application in which this
  situation occurs is the experimental study of fuel injection in
  automobile engines
  \cite{begg-2009-ER,kaplanski-2010-EJMB,dan-2012-JNM}: measurements
  in the injected two-phase spray do not always provide reliable
  velocity fields in the vortex bubble $\Omega_b$ because of the high
  density of seeding particles \cite{prosperi-2007}. It is then
  necessary to theoretically reconstruct parts of the flow field not
  accessible to measurements.  As demonstrated by the results of our
  test problem concerning Hill's vortex, even in such a restricted
  setting, the vorticity function can be accurately reconstructed with
  our approach. In our second example based on actual DNS data, we
  will show how the proposed approach can improve the accuracy of an
  inviscid vortex model derived from a purely empirical fit for the
  vorticity function. The predictions of our model will also be
  compared to classical reconstruction methods based on fitting
  theoretical vortex-ring models to the entire velocity field inside
  the vortex-ring bubble. We will show that our approach offers a good
  approximation of both the structure of the vorticity field and its
  integral characteristics, which is not the case with classical
  reconstruction methods. This is quite remarkable, since only partial
  information about the velocity field is used in our method.}

The structure of the paper is as follows: in the next {two sections}
we introduce the equations satisfied by the steady inviscid vortex
rings and formulate the reconstruction problem in terms of an
optimization approach. In Section \ref{sec:grad} we propose a
gradient-based solution method and derive the gradient formula.  The
computational algorithm is described in Section \ref{sec:comput},
together with the tests validating the method used for the computation
of the gradients.  The proposed method is first validated against
{a known analytical solution} (Hill's vortex) in Section
\ref{sec:results}.  The approach is then applied to a challenging
problem {of reconstructing an optimal vorticity function from
  realistic DNS data in Section \ref{sec:DNS}.}  Discussion and final
conclusion are deferred to Section \ref{sec:final}.

\section{Physical Problem and Governing Equations}
\label{sec:model}

In this section we present the equations satisfied by our vortex
model.  We consider incompressible axisymmetric vortex rings without
swirl {and add that formally a very similar description also
  holds for 2D inviscid flows}.  If a stationary solution is sought,
it is more convenient to describe the flow in the frame of reference
moving with the translation velocity $W\e_z$ (assumed constant) of the
vortex ring (see Figure \ref{fig-intro-dns}). A divergence-free
velocity field is {constructed} by defining the Stokes streamfunction
$\psi$ \cite{batchelor-1988,saffman-1992} such that
\begin{equation}
\label{eq-gen-vit-psi}
v_z=\frac{1}{r} \frac{\pl \psi}{\pl r},\quad   v_r=-\frac{1}{r}
\frac{\pl \psi}{\pl z}, \quad v_\theta=0.
\end{equation}
{The azimuthal component of the vorticity vector is then given
  by}
\begin{equation}
\label{eq-gen-om-th}
\omega = \frac{\pl v_r}{\pl z} - \frac{\pl v_z}{\pl r}.
\end{equation} 
Combining  \eqref{eq-gen-vit-psi}  and \eqref{eq-gen-om-th} results in
an elliptic partial differential equation for the streamfunction
\begin{equation} 
\L\psi = - \omega \quad \mbox{in}\,\, \Pi = \left\{ (z,r) \in \RR^2, r > 0 \right\},
\label{eq-gen-psiom}
\end{equation} 
where $\L$ is {defined as}
\begin{equation} 
\quad \L := \frac{\pl }{\pl z}\left( \frac{1}{r} \frac{\pl }{\pl z}\right) +  \frac{\pl }{\pl r}\left( \frac{1}{r} \frac{\pl }{\pl r}\right) =  \bnabla\cdot\left(\frac{1}{r} \bnabla\right), \quad \text{where} \ 
\bnabla := \left({\frac{\partial}{\pl z}},{\frac{\partial}{\pl
      r}}\right)^T.
\label{eq-gen-op-l}
\end{equation}  
The boundary condition {required for equation}
\eqref{eq-gen-psiom} accounts for an external flow around the vortex
which is uniform at infinity with velocity $-W\e_z$ (see Figure
\ref{fig-intro-dns})
\begin{equation} 
\Psi := \psi + \frac{1}{2}W r^2 \rightarrow 0 \quad  \mbox{as}\quad |\x| := \sqrt{z^2+r^2}
\rightarrow \infty.
\label{eq-psi-lab}
\end{equation}
We note that $\Psi$ is the Stokes streamfunction in the laboratory
frame of reference; $\Psi$ also satisfies the PDE
\eqref{eq-gen-psiom}, since $\L (k + \frac{1}{2}W r^2)=0$ for any
constants $W$ and $k$.

We recall that for inviscid and steady flows in the frame of reference
moving with the vortex ring, the transport equation for the vorticity
reduces to \eqref{eq-om-fpsi}. Problem \eqref{eq-gen-psiom} {can} be
reduced to a semi-linear elliptic {system by considering a particular
  form of the vorticity function as given, for example, in
  \eqref{eq-gen-frz} \cite{fraenkel-1974-AR}.} A different
reformulation of the problem, {namely,} as a semi-linear Dirichlet
boundary-value problem for the Laplacian {operator} in cylindrical
coordinates in $\RR^5$, was introduced in \cite{ni-1980}.  This made
possible the use of variational techniques to prove existence results
\cite{ni-1980,ambrosetti-1989}, symmetry \cite{esteban-1983} {and}
asymptotic behaviour \cite{tadie-1994-green} of solutions.

In the present study, we formulate the vortex-ring problem in the
domain {$\Omega \subset \RR^2$} defined as the cross-section of the
vortex bubble {$\Omega_b$} in the meridian half plane $r>0$ (see
Figure \ref{fig-intro-dns}b).  The domain $\Omega$ is then {bounded} by
the dividing streamline ($\psi=0$) {containing} the front ($A$)
and rear ($B$) stagnation points. On the axis of symmetry ($r=0$) the
radial velocity $v_r$ vanishes {which is consistent with the
  relation $\psi = 0$ holding there.}  These two parts of the boundary
of the vortex bubble will be denoted $\gamma_b$ and $\gamma_z$,
respectively.  Thus, the governing system for vortex rings takes the
final form
\begin{subequations}
\label{eq:Euler2D}
\begin{alignat}{2}
\L \psi & = - r \, f(\psi) 
&\quad & \textrm{in} \ \Omega, \label{eq:Euler2Da} \\
\psi & = 0 &  & \textrm{on} \ \gamma := \gamma_z \cup \gamma_b.
\label{eq:Euler2Db} 
\end{alignat}
\end{subequations}
Although the fore-and-aft symmetry is not enforced in solving system
\eqref{eq:Euler2D}, most (albeit not all) solutions of this system
obtained in our study will have this property. {Hereafter we will use
  as diagnostic quantities the following integral characteristics of
  the vortex rings: circulation $\Gamma$, impulse (in the horizontal
  direction $z$) $I$, and energy $E$. Using the vorticity function
  $f(\psi)$, they can be expressed \cite{saffman-1992} in terms of the
  following integrals over domain $\Omega$ (figure \ref{fig-intro-dns}b)
\begin{subequations}
\label{eq:GIE}
\begin{align}
\Gamma &:= \int_{\Omega} r f(\psi(z,r)) \, drdz, \label{eq:G} \\
I &:= \pi \int_{\Omega} r^3 f(\psi(z,r)) \, drdz, \label{eq:I} \\
E &:= \pi \int_{\Omega} r f(\psi(z,r)) \psi(z,r) \, drdz. \label{eq:E}
\end{align}
\end{subequations}}

\section{Formulation of the Reconstruction Problem}
\label{sec:formulation}

{In this section we formulate the reconstruction problem as an inverse
  problem of source identification amenable to solution using
  variational optimization techniques.}  Before we can precisely
{state this formulation}, we need to characterize the admissible
vorticity functions $f$. Their domain of definition will be restricted
to the interval $\I := [0,\psi_{\max}]$, where $\psi_{\max} > \max_{\x
  \in \Omega} \psi(\x)$ is chosen arbitrarily, so that $f \; : \; \I
\rightarrow \RR$.  We will refer to $\I$ as the ``identifiability
interval'' \cite{bvp10}.  Next, we note that in order to guarantee the
existence of nontrivial solutions to nonlinear elliptic boundary-value
problems of the type \eqref{eq:Euler2D}, the vorticity function
$f(\psi)$ must be positive (we refer the reader to the monographs
\cite{e02,c07} for a more detailed discussion of this issue).
Regarding the regularity of the vorticity function, {we will
  restrict our attention to continuous functions $f$ which is required
  due to certain technical aspects of the reconstruction algorithm
  (see Section \ref{sec:grad}\ref{sec:gradL2}). While this assumption
  does exclude vortex-ring models with vorticity support compact in
  $\Omega_b$, such as the NF model, cf.~relation \eqref{eq-gen-frz},
  such continuous vorticity functions are more appropriate for
  practical applications motivating this study. More specifically,} we
will assume that $f$ belongs to the Sobolev space $H^1(\I)$ of
continuous functions defined on $\I$ with square-integrable gradients.
The inner product defined in this space is
\begin{equation}
\forall_{z_1,z_2 \in H^1(\I)} \qquad
\big\langle z_1, z_2 \big\rangle_{{H^1(\I)}} = 
\int_0^{\psi_{\text{max}}} z_1 z_2 + {\ell^2} \Dpartial{z_1}{s} \Dpartial{z_2}{s}\, ds,
\label{eq:ipH1}
\end{equation}
where $\ell \in \RR^+$ is a parameter with the meaning of a ``length
scale'' {(the significance of this parameter will be discussed
  further in Section \ref{sec:grad}\ref{sec:sobolev})}. We can now
state the reconstruction problem as follows
\begin{problem}
  Given the measurements $m \; : \; \gamma_b \cup \gamma_z \rightarrow
  \RR$ of the velocity component tangential to the boundaries
  $\gamma_b$ and $\gamma_z$, find a vorticity function $\hat{f} \in
  H^1(\I)$, such that the corresponding solution of \eqref{eq:Euler2D}
  matches data $m$ as well as possible in the least-squares sense.
\label{P1}
\end{problem}
For the purpose of the numerical solution, we will recast Problem
\ref{P1} as a variational optimization problem which can be solved
using a suitable gradient-based method. Since the tangential velocity
at the boundary is ${\vec{v}}\cdot \n^\perp =
\frac{1}{r}\Dpartial{\psi}{n}$, we define a cost functional $\J \; :
\; H^1(\I) \rightarrow \RR$ as
\begin{equation}
\J(f) := \frac{\alpha_b}{2} \int_{\gamma_b} 
\left( \frac{1}{r}\Dpartial{\psi}{n}\bigg|_{\gamma_b} - m \right)^2 \, d\sigma +
\frac{\alpha_z}{2} \int_{\gamma_z} 
\left( \frac{1}{r}\Dpartial{\psi}{n}\bigg|_{\gamma_z} - m \right)^2 \, d\sigma,
\label{eq:J}
\end{equation}
where $\alpha_b$ and $\alpha_z$ assume the values $\{0,1\}$ depending
on {which part(s)} of the domain boundary the measurements are
available on.  {At this point we remark that measurement data
  distributed over some finite-area region $\R \in \Omega$ can also be
  used and in such case the line integrals in \eqref{eq:J} will be
  replaced with suitable area integrals over $\R$.}  The optimal
reconstruction $\hat{f}$ will thus be obtained via solution of the
following minimization problem
\begin{equation}
\hat{f} := \argmin_{f \in H^1(\I)} \ \J(f),
\label{eq:min}
\end{equation}
{where ``$\argmin$'' denotes the argument minimizing the
  objective function.}  In some situations it {may} be necessary
to enforce the nonnegativity $f(\psi) \ge 0$ of the vorticity function
(functions obtained by imposing this property will be denoted $f_+$).
Rather than including an inequality constraint in optimization problem
\eqref{eq:min}, this can be achieved in a straightforward manner by
expressing $f_+ = (1/2) g^2$, where $g$ is a real-valued function
defined on $\I$, and then recasting problem \eqref{eq:min} in terms of
the new function $g$ as {the} control variable.

Problem \ref{P1} is an example of an inverse problem of source
{identification}. However, in contrast to the most common
problems of this type \cite{i98}, in which the source function depends
on the independent variables (e.g., on $\x$), in Problem \ref{P1} the
source $f$ is sought as a function of the {\em state variable} $\psi$.
As will be shown in the following section, to address this aspect of
the problem, a specialized version of the adjoint-based gradient
approach will be developed.

\section{Gradient-Based Solution Approach}
\label{sec:grad}

In this section we first describe the general optimization formulation
which is followed by the derivation of a convenient expression for the
cost functional gradient. Finally, we discuss the calculation of
smoothed Sobolev gradients.

\subsection{Minimization Algorithm}
\label{sec:min}
 
For simplicity, the solution approach to Problem \ref{P1} we present
below will not address the positivity constraint which {can be
  accounted for in a straightforward manner using the substitution
  mentioned at the end of the previous section.}  Solutions to problem
\eqref{eq:min} are characterized by the following first-order
optimality condition
\begin{equation}
\forall_{f' \in H^1(\I)} \quad \J'(\hat{f}; f') = 0,
\label{eq:dJ0}
\end{equation}
where the G\^ateaux differential $\J'(f; f') :=
\lim_{\epsilon\rightarrow 0} \epsilon^{-1} \left[ \J(f+\epsilon f') -
  \J(f)\right]$ of functional \eqref{eq:J} is
\begin{equation}
\J'(f; f') = \alpha_b \int_{\gamma_b} \left(  \frac{1}{r}
\Dpartial{\psi}{n}\bigg|_{\gamma_b} - m \right) \, 
 \frac{1}{r} \Dpartial{\psi'}{n}\bigg|_{\gamma_b} d\sigma +
\alpha_z \int_{\gamma_z} \left(  \frac{1}{r}
\Dpartial{\psi}{n}\bigg|_{\gamma_z} - m \right) \, 
 \frac{1}{r} \Dpartial{\psi'}{n}\bigg|_{\gamma_z} d\sigma,
\label{eq:dJ}
\end{equation}
in which the variable $\psi'$ satisfies the linear perturbation
equation
\begin{subequations}
\label{eq:dEuler2D}
\begin{alignat}{2}
\bnabla \cdot\left(\frac{1}{r} \bnabla \psi'\right) 
+ r f_{\psi}(\psi)\, \psi' & = - r \, f' 
&\quad & \textrm{in} \ \Omega, \label{eq:dEuler2Da} \\
\psi' & = 0 &  & \textrm{on} \ \gamma, 
\label{eq:dEuler2Db}
\end{alignat}
\end{subequations}
where $f_{\psi} := \dfrac{df}{d\psi}$ and $f'$ is the ``direction'' in
which the differential is computed in \eqref{eq:dJ}. {The
  presence of the derivative $f_\psi$ in \eqref{eq:dEuler2Da} is the
  reason explaining the regularity requirements imposed on $f$
  (cf.~Section \ref{sec:grad}).}

The optimal reconstruction can be obtained as $\hat{f} =
\lim_{k\rightarrow \infty} f^{(k)}$, where the approximations
$f^{(k)}$ can be computed with the following gradient descent
algorithm
\begin{equation}
\begin{aligned}
f^{(k+1)} & =  f^{(k)} - \tau_k \nabla\J(f^{(k)}), \quad k=1,2,\dots \\ 
f^{(1)} & =  f_0,
\end{aligned}
\label{eq:desc}
\end{equation}
in which $f_0$ is the initial guess and $\tau_k$ represents the length
of the step along the descent direction at the $k$-th iteration. For
the sake of simplicity, formulation \eqref{eq:desc} corresponds to the
steepest-descent algorithm, however, {in actual computations} we shall
prefer more advanced minimization techniques, such as the conjugate
gradient method \cite{nw00} (see Section \ref{sec:comput}). {We
  note that optimality condition \eqref{eq:dJ0} and the associated
  gradient descent \eqref{eq:desc} characterize only {\em local}
  minimizers and establishing a priori whether a given minimizer is
  global is not possible. This is a consequence of the nonconvexity of
  Problem \ref{P1} resulting from the nonlinearity of governing system
  \eqref{eq:Euler2D}. Global maximizers are sought by solving problem
  \eqref{eq:desc} repeatedly using a range of different initial
  guesses $f_0$.}

\subsection{Derivation of the Gradient {Expression}}
\label{sec:gradL2}

A key element of descent algorithm \eqref{eq:desc}
is the cost functional gradient $\nabla\J(f)$. Assuming that G\^ateaux
differential \eqref{eq:dJ} is a bounded linear functional defined on a
Hilbert space $\X$ (e.g., $\X = L_2(\I)$ or $\X = H^1(\I)$), i.e.,
$\J'(f;\cdot) \; : \; \X \rightarrow \RR$, an expression for the
gradient $\nabla^{\X} \J(f)$ can be obtained from \eqref{eq:dJ} employing
the Riesz representation theorem \cite{l69}
\begin{equation}
\J'(f; f') = \Big\langle \nabla^{\X} \J(f), f' \Big\rangle_{\X},
\label{eq:Riesz}
\end{equation}
with $\langle . , .\rangle_{\X}$ denoting the inner product in
{the} space $\X$.  We note that representation \eqref{eq:dJ} is
not yet consistent with \eqref{eq:Riesz}, since the perturbation $f'$
is not explicitly present in it, but is instead hidden in the source
term of perturbation equation \eqref{eq:dEuler2Da}. In order to
identify an expression for the gradient consistent with
\eqref{eq:Riesz}, we introduce the {\em adjoint} variable $\psi^* \; :
\; \Omega \rightarrow \RR$. Integrating \eqref{eq:dEuler2Da} against
$\psi^*$ over $\Omega$ and then integrating by parts {twice} we obtain
\begin{equation}
\begin{aligned}
0 = & \int_{\Omega} \psi^*\, \left[ \bnabla \cdot\left(\frac{1}{r} \bnabla \psi'\right) 
+ r f_{\psi}(\psi)\, \psi' \right]\, d\Omega + \int_{\Omega} \psi^*\, r \, f' d\Omega \\
= & \int_{\Omega} \psi'\, \left[ \bnabla \cdot \left(\frac{1}{r} \bnabla \psi^*\right) 
+ r f_{\psi}(\psi)\, \psi^* \right]\, d\Omega + \int_{\Omega} \psi^*\, r \, f' d\Omega \\
& + \int_{\gamma_b \cup \gamma_z}  \frac{1}{r}
\left(\psi^* \Dpartial{\psi'}{n} - \psi' \Dpartial{\psi^*}{n}\right)\,d\sigma.
\end{aligned}
\label{eq:I1}
\end{equation}
Using boundary condition \eqref{eq:dEuler2Db} and defining the {\em
  adjoint system} as follows
\begin{subequations}
\label{eq:aEuler2D}
\begin{alignat}{2}
\bnabla \cdot\left(\frac{1}{r} \bnabla \psi^*\right) 
+ r f_{\psi}(\psi)\, \psi^* & =  0
&\quad & \textrm{in} \ \Omega, \label{eq:aEuler2Da} \\
\psi^* & =   \alpha_b \left(\frac{1}{r}\Dpartial{\psi}{n}\bigg|_{\gamma_b} - m\right) &  & \textrm{on} \ \gamma_b, 
\label{eq:aEuler2Db} \\
\psi^* & = \alpha_z \left(\frac{1}{r}\Dpartial{\psi}{n}\bigg|_{\gamma_z} - m\right) &  & \textrm{on} \ \gamma_z, 
\label{eq:aEuler2Dc} 
\end{alignat}
\end{subequations}
identity \eqref{eq:I1} simplifies to
\begin{equation}
  \J'(f; f') = - \int_{\Omega} \psi^*\, r \, f' d\Omega.
\label{eq:I2}
\end{equation}
Although perturbation $f'$ appears explicitly in \eqref{eq:I2}, this
expression still is not in a form consistent with Riesz representation
\eqref{eq:Riesz}, because the latter requires an inner product with
$s$ (equivalently, $\psi$) as the integration variable.  We address
this issue by expressing $f'(\psi)$ in terms of the following integral
transform
\begin{equation}
  f'(\psi(\x)) = \int_0^{\psi_{\max}} \delta(\psi(\x) - s) f'(s)\, ds,
\quad \x \in \Omega,
\label{eq:fx}
\end{equation}
where $\delta(\cdot)$ is the Dirac delta distribution. Plugging
\eqref{eq:fx} into \eqref{eq:I2} and then using Fubini's theorem to
{exchange} the order of integration, we obtain
\begin{equation}
\begin{aligned}
  \J'(f; f') & = - \int_{\Omega} \psi^*\, r \, 
\left[ \int_0^{\psi_{\max}} \delta(\psi(\x) - s) f'(s)\, ds \right] d\Omega \\
& = - \int_0^{\psi_{\max}} f'(s) \left[ 
\int_{\Omega} \psi^*\, r \,\delta(\psi(\x) - s)\, d\Omega\right] \, ds
\end{aligned}
\label{eq:I3}
\end{equation}
which is already consistent with Riesz representation
\eqref{eq:Riesz}. Although this is not the gradient used in our actual
calculations, we first identify the $L_2$ gradient of $\J$ and then
obtain from it the required Sobolev gradient $\nabla^{H^1}\J$ as shown
in Section \ref{sec:grad}\ref{sec:sobolev}. Thus, setting $\X =
L_2(\I)$, relation \eqref{eq:Riesz} becomes $\J'(f; f') = \int_{\I}
\nabla\J(s) f'(s) \,ds$ which, together with \eqref{eq:I3}, yields
\begin{equation}
\nabla^{L_2}\J(s) = -\int_{\Omega} \psi^*\, r \,\delta(\psi(\x) - s)\, d\Omega
= - \int_{\gamma_s} \psi^*\, r \, \left(\Dpartial{\psi}{n}\right)^{-1} \, d\sigma,
\quad s \in [0,\psi_{\max}].
\label{eq:gradL2}
\end{equation}
The expression on the right-hand side (RHS) of \eqref{eq:gradL2} shows
that, for a given $s\in\I$, the gradient $\nabla^{L_2}\J(s)$ can be
evaluated as a contour integral on the level set
\begin{equation}
\gamma_s := \{ \x \in \Omega \, : \,  \psi(\x) = s\}.
\label{eq:Cs}
\end{equation}

{We add that an essentially identical approach will remain
  applicable when the measurements are available over a finite-area
  region $\R$ rather than on the contours $\gamma_b$ and $\gamma_z$.
  The only difference is that the adjoint system will be ``forced''
  through a source term (with the support equal to $\R$) on the RHS of
  \eqref{eq:aEuler2Da}, instead of through boundary conditions
  \eqref{eq:aEuler2Db}--\eqref{eq:aEuler2Dc} as discussed above.}

\subsection{Sobolev Gradients}
\label{sec:sobolev}

We now proceed to discuss how Sobolev gradients $\nabla\J =
\nabla^{H^1}\J$ employed in gradient-descent approach \eqref{eq:desc}
can be derived from the $L_2$ gradients obtained in \eqref{eq:gradL2}.
{We remark that this additional regularity is required for the
  consistency of the entire approach, since the reconstruction with
  $L_2$ gradients would not guarantee (weak) differentiability of the
  vorticity function $f(\psi)$, thus rendering adjoint system
  \eqref{eq:aEuler2D} ill-posed (because of the term $f_{\psi}$ in
  \eqref{eq:aEuler2Da}).}  This will be done using inner product
\eqref{eq:ipH1} in Riesz identity \eqref{eq:Riesz}. In addition to
enforcing smoothness of the reconstructed vorticity functions, this
formulation also allows us to impose the desired behavior at the
endpoints of interval $\I$ via suitable boundary conditions {(we refer
  the reader to \cite{bvp10} for a more in-depth discussion of these
  issues)}. As regards the behavior of the gradients $\nabla^{H^1} \J$
at the endpoints of interval $\I$, we can require the vanishing of
either the gradient itself or its derivative with respect to $s$. In
the present study we prescribe the homogeneous Neumann boundary
condition at the right endpoint of the identifiability interval $\I$
\begin{equation}
\frac{d}{ds}(\nabla^{H^1} \J) = 0 \quad \textrm{at} \ s=\psi_{\text{max}}
\label{eq:gradH1BCr}
\end{equation}
which implies that, with respect to the initial guess $f_0$, {at
  $s=\psi_{\text{max}}$} iterations \eqref{eq:desc} can modify the
values, but not the slope, of the reconstructed functions $f^{(k)}$.
As regards the behavior of the Sobolev gradients at the left endpoint,
we will consider either Dirichlet or Neumann boundary conditions
\begin{subequations}
\label{eq:gradJH1l}
\begin{alignat}{2}
\nabla^{H^1} \J & = 0 \quad && \textrm{at} \ s=0, \label{eq:gradJH1ld} \\
\frac{d}{ds}(\nabla^{H^1} \J) & = 0 \ && \textrm{at} \ s=0 \label{eq:gradJH1ln}
\end{alignat}
\end{subequations}
which will preserve, respectively, the value or the slope of the
initial guess $f_0$ at $s=0$.  We refer the reader to \cite{pnm14} for
a discussion of other possible choices of boundary conditions imposed
on the Sobolev gradients in an identification problem with a similar
structure. {We emphasize that the choice of the boundary behavior
  of the Sobolev gradients plays in fact a significant role from the
  physical point of view. Together with the behaviour of the initial
  guess $f_0$ in the neighbourhood of $s=0$ and $s=\psi_{\text{max}}$,
  it expresses our hypotheses on the properties of the optimal
  reconstruction $\hat{f}(\psi)$ for the limiting values of $\psi$
  where no measurement data is available. The need to supplement
  measurement data with some auxiliary information about the solution
  is quite typical for inverse problems \cite{t05}. Additional comments
  about the specific physical meaning of the boundary conditions
  imposed on the Sobolev gradients will be provided in Sections
  \ref{sec:results} and \ref{sec:DNS}.}

Identifying expression \eqref{eq:Riesz} in which $\X = H^1(\I)$ with
the inner product given in \eqref{eq:ipH1}, integrating by parts and
using boundary conditions \eqref{eq:gradH1BCr}--\eqref{eq:gradJH1l} we
obtain the following elliptic boundary-value problem {on $\I$}
defining the Sobolev gradient $\nabla^{H^1} \J$
\begin{subequations}
\label{eq:gradJH1}
\begin{alignat}{2}
{\left( I - {\ell}^2 \frac{d^2}{ds^2} \right)} \nabla^{H^1} \J &= \nabla^{L_2} \J & \qquad & \textrm{in} \ \I, \label{eq:gradJH1_1} \\
\left.\begin{aligned}
\nabla^{H^1} \J \quad & \\
\frac{d}{ds} \nabla^{H^1} \J \quad &
\end{aligned}\right\} &= 0 & & \textrm{at} \ s = 0, \label{eq:gradJH1_2} \\
\frac{d}{ds} \nabla^{H^1} \J &= 0 & & \textrm{at} \ s = \psi_{\text{max}}, \label{eq:gradJH1_3} 
\end{alignat}
\end{subequations}
where the expression for $\nabla^{L_2} \J$ is given in
\eqref{eq:gradL2}. A slightly different way of obtaining Sobolev
gradients in identification problems with analogous structure is
discussed in \cite{bvp10}.

It is well known \cite{pbh04} that extraction of cost functional
gradients in the space $H^1$ with the inner product defined as in
\eqref{eq:ipH1} can be regarded as low-pass filtering of $L_2$
gradients with the cut-off wavenumber given by {$\ell^{-1}$. The
  quantity $\ell$ admits a clear physical meaning as the smallest
  ``length-scale'' (with the magnitude of the streamfunction $\psi$
  playing the role of ``length'') which is retained when the Sobolev
  gradient is extracted according to \eqref{eq:gradJH1}. In other
  words, features of the $L_2$ sensitivity \eqref{eq:gradL2} with
  characteristic length scales smaller than $\ell$ are removed during
  gradient preconditioning. Therefore, by choosing $\ell$ to represent
  the characteristic variation of the streamfunction $\psi$ in the
  problem, this mechanism allows us to eliminate in a controlled
  manner undesired small-scale components which may be present in the
  $L_2$ gradients due to noise in the measurements, numerical
  approximation errors, etc.} One approach which has been found to
work particularly well \cite{pbh04} is to start with a relatively
large value of $\ell$, which gives smooth gradients suitable for
reconstructing large-scale features of the solution, and then decrease
it with iterations, which allows one to zoom in on
{progressively} smaller features of the solution. This is the
approach we adopt here by setting $\ell^{(k)}$, the value of the
length-scale used {in \eqref{eq:gradJH1}} at the $k$-th
iteration, as
\begin{equation}
\ell^{(k)} = \zeta^k \, \ell^{(0)}, \quad k>0,
\label{eq:zeta}
\end{equation}
where $ \ell^{(0)}$ is some initial value and $0 < \zeta < 1$
{the decrement factor}.

\section{Computational Algorithm}
\label{sec:comput}

As is evident from Section \ref{sec:formulation}, the reconstruction
algorithm requires the solution of several linear and nonlinear
elliptic boundary-value problems in one or two spatial dimensions,
namely, the governing system \eqref{eq:Euler2D}, the adjoint system
\eqref{eq:aEuler2D} and the preconditioning system for the Sobolev
gradients \eqref{eq:gradJH1}. In addition, evaluation for the $L_2$
gradients is somewhat involved, because the integrals in
\eqref{eq:gradL2} are evaluated on the level sets $\gamma_s$ which
have to be identified, cf.~definition \eqref{eq:Cs}. All of these
technical issues were easily handled using the {freely available}
finite-element software {\tt FreeFem++} \cite{freefem,hecht-2012-JNM}.
This generic PDE solver offers the possibility of using a large
variety of triangular finite elements with an integrated grid
generator in two or three dimensions.  {\tt FreeFem++} {is equipped}
with its own high-level programming language with syntax close to
mathematical formulations.  It was recently used to solve different
types of partial differential equations, \eg {Schr{\"o}dinger and
  Gross-Pitaevskii equations} \cite{dan-2010-SISC,dan-2010-JCP},
incompressible Navier-Stokes equations \cite{bp11a}, {Poisson}
equations with nonlinear source terms \cite{dan-2013-AMM} and
Navier-Stokes-Boussinesq equations \cite{dan-2014-JCP}.  The main
advantage of {employing} {\tt FreeFem++} for {the present}
problem {is} the simplicity in using different finite-element meshes
for each sub-problem making the interpolation or computation of
integrals very easy and accurate.  {Below we briefly describe the
  implementation of key elements of the computational algorithm which
  are then validated in the following section}.

\subsection{Main Computational Modules} 
\label{sec:modules}

The computational algorithm consists {of} the following main
modules:

$\bullet$ [{Definition of the mesh and the} associated finite-element
spaces] We define here the boundaries $\gamma_b$ and $\gamma_z$ and
build a triangular mesh {covering} the vortex domain $\Omega$ (see
Figure \ref{fig:algo-lset}). The mesh density is characterized by
$N_x$ representing the number of segments per unit length in the
discretization of the domain boundaries. The finite element space
$V_h$ is defined {such that all {dependent} variables are
  represented using piecewise quadratic $P^2$ finite elements. Cost
  function \eqref{eq:J} is computed with a 6-th order Gauss quadrature
  formula.

  $\bullet$ [One-dimensional interpolation] The vorticity function
  $f(s)$ is tabulated {at $N_f$ discrete values $s_i \in
    [0,\psi_{\max}]$, $i=1,\dots,N_f$}. The value $\psi_{\max}$
  {(cf.~Section \ref{sec:formulation})} is set depending on {a
    particular} reconstruction case. To obtain values {of} $f$
  and its derivative $f_\psi$ for non-tabulated values {of $\psi$} we
  use cubic spline interpolation.

$\bullet$ [{Solution of} direct problem \eqref{eq:Euler2D}]
{Given the nonlinearity of this problem, we use Newton's method
  with $p$-th} iteration consisting in computing the solution $q :=
(\psi^p - \psi^{p+1})$ of the following variational problem
\begin{equation}
\int_{\Omega } \frac{1}{r} \bnabla q \cdot\bnabla v \, d\Omega - \int_{\Omega } {r f_\psi(\psi^p)} q v \, d\Omega= 
\int_{\Omega } \frac{1}{r} \bnabla \psi^p \cdot\bnabla v \, d\Omega- \int_{\Omega } {r f(\psi^p)}  v \, d\Omega, \quad \forall v \in V_h. 
\label{eq:algo-newton}
\end{equation}
This problem is solved {efficiently} in {\tt FreeFem++} by
building the corresponding matrices (spline interpolation is used
{to evaluate $f(\psi(\x))$ and $f_\psi(\psi(\x))$}). Newton's
iterations are stopped when $\|q\|_2 \leq \varepsilon_N$ with
$\varepsilon_N = 10^{-6}$.

$\bullet$ [{Solution of} adjoint problem \eqref{eq:aEuler2D}]
{Given the linearity of this problem}, this consists in solving
the weak formulation
\begin{equation}
\int_{\Omega } -\frac{1}{r} \bnabla \psi^* \bnabla v \, d\Omega + \int_{\Omega } {r f_\psi(\psi)} \psi^* v \, d\Omega = 0, \quad \forall v \in V_h, 
\label{eq:algo-adj}
\end{equation}
with Dirichlet boundary conditions
\eqref{eq:aEuler2Db}-\eqref{eq:aEuler2Dc}, {which} takes two
lines of code in {\tt FreeFem++}.

$\bullet$ [{Computation of} $L_2$ gradient] To use formula
\eqref{eq:gradL2} for the $L_2$ gradient {$\nabla^{L_2}\J(s)$},
for each value {$s_i$, $i=1,\dots,N_f$,} in the table defining
the {discretized} vorticity function {$f(s_i)$}, we
{construct} the corresponding level set {$\gamma_{s_i}$} and
mesh {its interior} (see Figure \ref{fig:algo-lset}). {The
  values of} $\psi^*$ and $\psi$ are $P^2$ interpolated on the new
mesh and the integral in \eqref{eq:gradL2} is then computed with a
6-th order Gauss quadrature formula.

$\bullet$ [{Computation of} $H^1$ gradient] To {obtain} the
$H^1$ gradient from the $L_2$ gradient we solve the one-dimensional
{boundary-value problem} \eqref{eq:gradJH1} with either
\eqref{eq:gradJH1_2} or \eqref{eq:gradJH1_3} {as the} boundary
condition. {This is a standard problem which can be solved in a
  straightforward manner using $P^1$ piecewise linear finite elements
  or second-order accurate centered finite differences.}

$\bullet$ [{Minimization algorithm}] With the cost functional
gradient evaluated as described above, we approximate the optimal
{vorticity} function $\hat{f}$ using the Polak-Ribiere variant of
the conjugate gradients algorithm \cite{nw00} which is an improved
version of descent algorithm \eqref{eq:desc}. The length of the step
$\tau_k$ at every iteration $k$ is determined by solving a line
minimization problem
\begin{equation}
\tau_k = \underset{\tau > 0}{\argmin} \;\J(f^{(k)} - 
\tau \, \bnabla \J(f^{(k)}))
\label{eq:linemin}
\end{equation}
using Brent's method \cite{pftv86}. 

Clearly, accurate evaluation of the cost functional gradients $\bnabla
\J(f)$ is a key element of the proposed reconstruction approach and
these calculations are thoroughly validated in the following section.

\begin{figure}
\begin{center}
\includegraphics[width=0.7\textwidth]{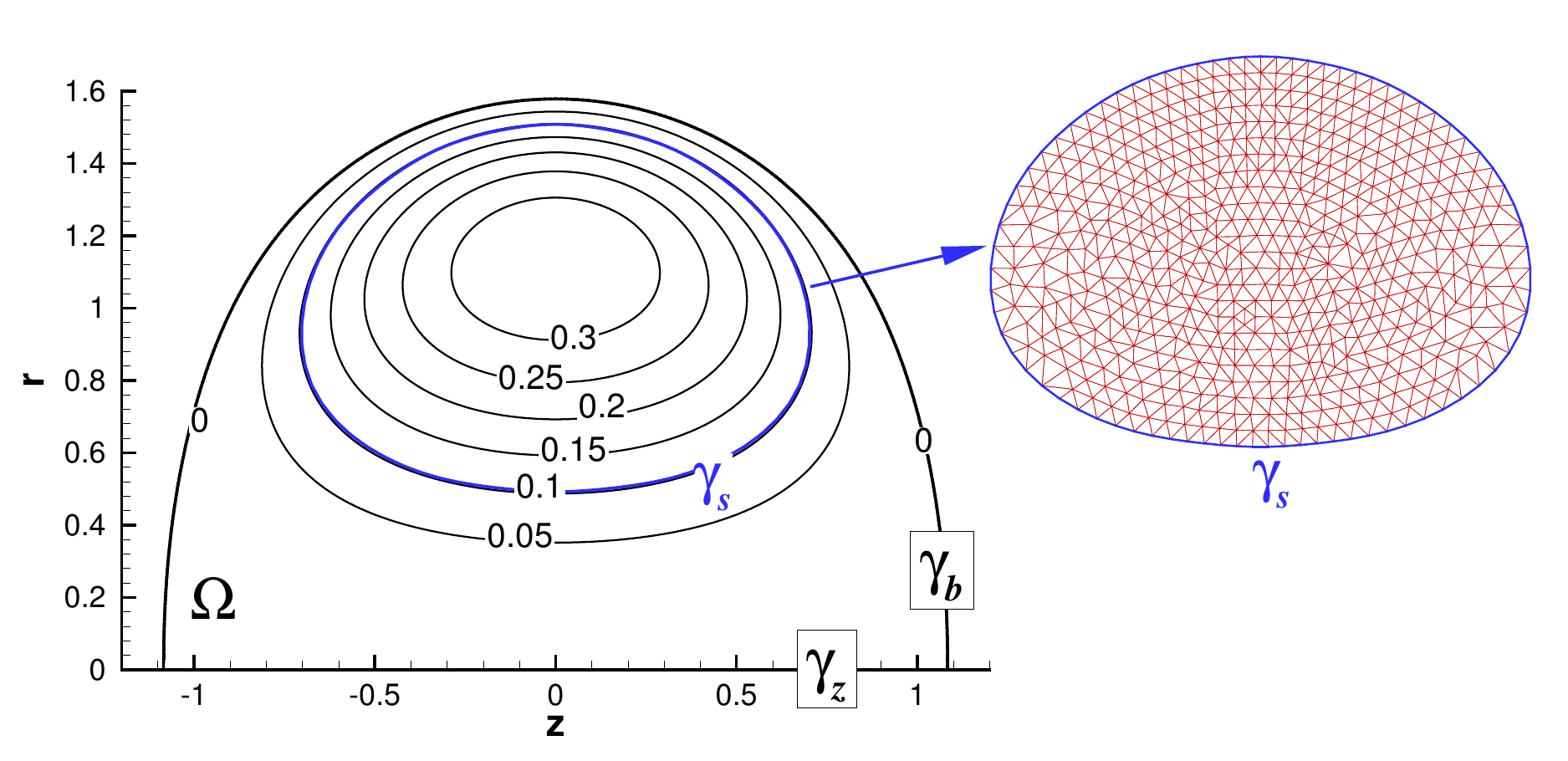}
\vspace*{-0.35cm}
\caption{{Schematic of the calculation of the $L_2$ gradient of
    the cost functional based on formula \eqref{eq:gradL2}. The level
    set $\gamma_s$ corresponding to $s=0.1$ is marked with a thick
    line, whereas the inset represents the mesh discretizing the
    domain bounded by $\gamma_s$.}}
\label{fig:algo-lset}
\end{center}
\end{figure}

\subsection{Validation of Cost Functional Gradients}
\label{sec:kappa}

In this section we analyze the consistency of the gradient $\bnabla
\J$ evaluated based on formula \eqref{eq:gradL2} with respect to
refinement of the two key numerical parameters in the problem, namely,
{$N_x$ and $N_f$ (see the previous section for definitions)}.
A standard test \cite{hnl02} consists in computing the G\^{a}teaux
differential $\J'(f; f')$ in some arbitrary direction $f'$ using
relations \eqref{eq:I3}--\eqref{eq:gradL2} and comparing it to the
result obtained with a forward finite--difference formula. Thus,
deviation of the quantity
\begin{equation}
\kappa(\epsilon) :=  \frac{  \epsilon^{-1} \left[\J(\psi_b+\epsilon\psi'_b) -
\J(\psi_b)\right]}{\int_0^{\psi_{\max}} f'(s) \bnabla\J(s) \, ds}
\label{eq:kappa}
\end{equation}
from the unity is a measure of the error in computing $\J'(f; f')$ (we
note that, in the light of identity \eqref{eq:Riesz}, expression in
the denominator of \eqref{eq:kappa} may be based on the $L_2$
gradients).  

The dependence of the quantity $\log|\kappa(\epsilon) -
1|$, which captures the number of significant digits of accuracy
achieved in the evaluation of \eqref{eq:kappa}, on $\epsilon$ is shown
in Figures \ref{fig:kappa}a and \ref{fig:kappa}b, respectively, for
increasing $N_f$ and $N_x$ while keeping the other parameter fixed.
\begin{figure}
\begin{center}
\includegraphics[width=0.8\textwidth]{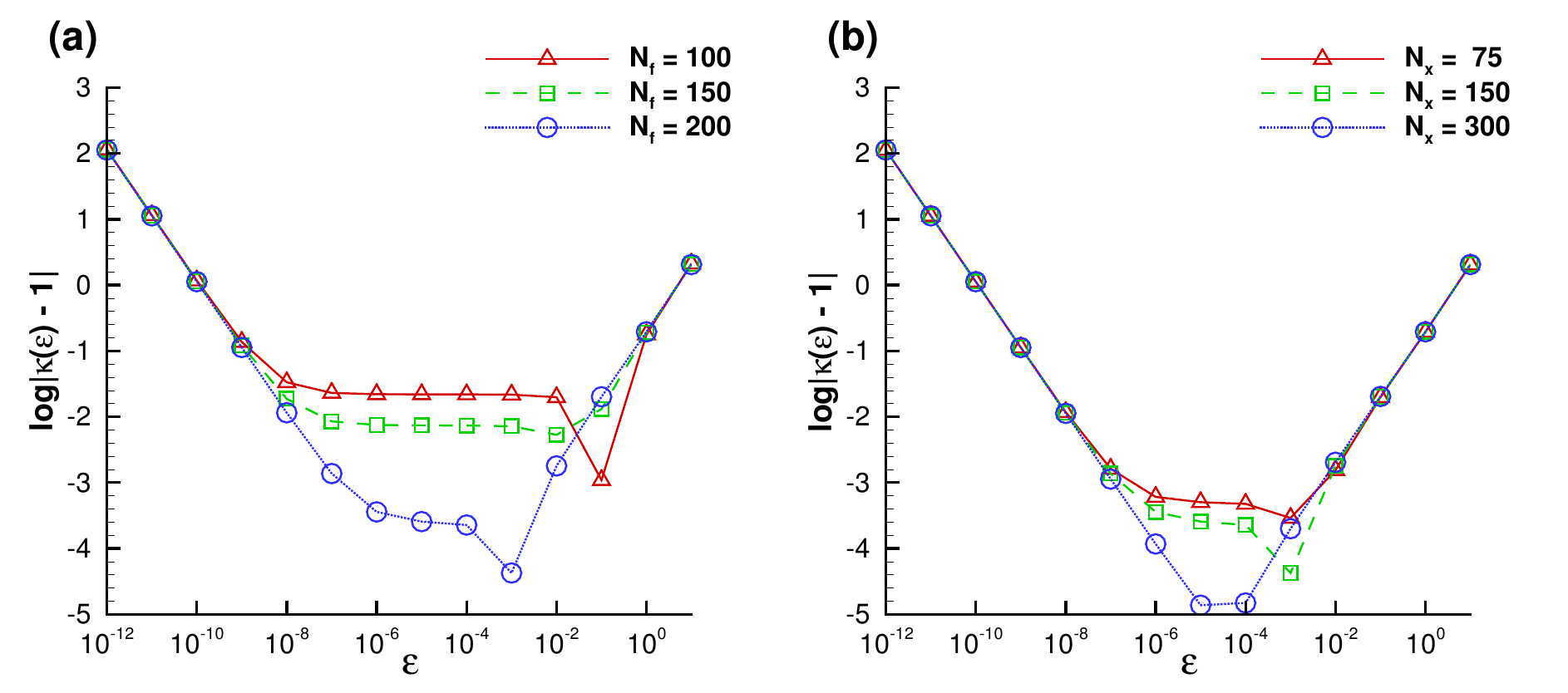}
\vspace*{-0.3cm}
\caption{Dependence of the diagnostic quantity $\kappa(\epsilon)$
  defined in \eqref{eq:kappa} on $\epsilon$ for (a) different
  discretizations of the identifiability interval $\I$ given by
  $N_f=100,150,200$ with $N_x=150$ fixed, and (b) different
  discretizations of the domain $\Omega$ given by $N_x=75,150,300$
  with $N_f=200$ fixed.}
\label{fig:kappa}
\end{center}
\end{figure}
These results were obtained in a configuration representing Hill's
vortex in which $C = 1/2$ and $\Omega$ is a half-circle of radius
$a=2$ ({see {Section \ref{sec:results}} for a precise
  definition of this test problem}), and some generic forms of the
reference vorticity function $f$ and its perturbation $f'$ were used.
As is evident from Figures \ref{fig:kappa}a and \ref{fig:kappa}b, the
values of $\kappa(\epsilon)$ approach the unity for $\epsilon$ ranging
over approximately 7 orders of magnitude as the discretization is
refined (i.e., as $N_f$ and $N_x$ increase).  We emphasize that, since
we are using the ``differentiate--then--discretize'' rather than
``discretize--then--differentiate'' approach, the gradient should not
be expected to be accurate up to the machine precision \cite{g03}.
{This is because of the presence of small, but nonzero, errors in
  the approximation of the different PDEs and the gradient expression
  \eqref{eq:gradL2}.}  The deviation of $\kappa(\epsilon)$ from the
unity for very small values of $\epsilon$ is due to the arithmetic
round--off errors, whereas for the large values of $\epsilon$ it is
due to the truncation errors, both of which are well known effects
\cite{hnl02}.  {In particular, the former effect is typical of
  all finite-difference techniques and as such is an artifact of
  formula \eqref{eq:kappa} usually employed to test the accuracy of
  adjoint-based gradient expressions.}  These results thus demonstrate
high accuracy of the computed gradients {and confirm that this
  accuracy can be systematically improved by refining the
  discretization.}

The {computational} results presented in next {two sections} were
obtained with the numerical resolution $N_x = 75$, corresponding to
$N_e=\O(10^4-10^5)$ finite elements discretizing domain $\Omega$ (the
exact number {varied depending on the specific test} problem), and
$N_f = 100$. \revt{Solution of optimization problem
  \eqref{eq:J}--\eqref{eq:min} typically requires $\O(1)$--$\O(10)$
  iterations terminated when $\|\nabla^{H^1}\J\|_{L_2(\I)}$ drops
  below $10^{-5}$. The costliest element of each iteration is solution
  of the line minimization problem \eqref{eq:linemin} which on average
  necessitates $\O(10)$ solutions of Euler system \eqref{eq:dEuler2D}.
  Overall, the computational time required for a single iteration
  using the resolutions mentioned above on a workstation with Intel i7
  processors is $\O(1)$ minutes with a rather modest memory
  footprint.}

\section{{Reconstruction of Inviscid Vortex Rings --- Hill's
    Spherical Vortex}}
\label{sec:results}

In this section we employ algorithm \eqref{eq:desc} to reconstruct the
vorticity function $f$ in {a test case involving} Hill's
spherical vortex {in which the exact form of $f$ is known.}
Then, in Section \ref{sec:DNS}, we will use our approach to
reconstruct the vorticity function $f(\psi)$ in a steady Euler flow
assumed to model an actual high-Reynolds number flow with concentrated
vortex rings.  Data for this reconstruction will be obtained from a
DNS of such a flow.

Hill's spherical vortex is a well known
\cite{batchelor-1988,saffman-1992} closed-form solution for which the
vortex bubble $\Omega$ is a sphere of radius $a$ and the vorticity
function is constant everywhere in $\Omega$, i.e.,
\begin{equation}
f(\psi) = C, \qquad C > 0, \quad \forall\, \psi(\x), \quad \x \in \Omega.
\label{eq:FHill}
\end{equation}
The flow outside the bubble approaches the uniform flow $W\e_z$ as
$|\x| \rightarrow \infty$. By matching the solution inside the bubble
with the {exterior} solution, the continuity of $\psi$ and
$\bnabla \psi$ on $\gamma$ gives the compatibility relationship
\begin{equation}
 W = \dfrac{2}{15} C  a^2.
 \label{eq-gen-hill-U}
\end{equation}
The complete {expression for the streamfunction in Hill's vortex
  can be found for example in
  \cite{lamb-1932,batchelor-1988,saffman-1992}.} The circulation,
impulse and energy then take the following values, cf.~\eqref{eq:GIE},
\begin{equation}
\Gamma_{\text{Hill}} = \frac{2}{3} C a^3, \quad
I_{\text{Hill}} = \frac{4}{15} Ca^5 \pi, \quad
E_{\text{Hill}} = \frac{4}{525} C^2 a^7 \pi.
\label{eq:GIEHill}
\end{equation}

It is interesting to note that Hill's vortex is not only an Euler
solution, but also satisfies the Navier-Stokes equation (in this
sense, it is related to the ``controllable flows'' introduced by
Truesdell \cite{ma77}). Indeed, if an additional pressure $-2C\mu z$
is included inside the bubble to balance the viscous term $\mu \Delta
\vec{v}=-2C\mu \e_z$, the Navier-Stokes equation is satisfied both
inside and outside the vortex \cite{saffman-1992}.  However, at the
boundary of the vortex ring, only the continuity of the velocity is
satisfied. The normal and tangential stresses are not continuous
across the boundary, therefore {Hill's vortex} is not an exact
solution of the complete Navier-Stokes system.

{In the test problem analyzed here we consider Hill's vortex in
  which without the loss of generality we set $a=2$ and $C=1/2$.}  We
assume that the measurements $m$ of the tangential velocity component
are available on the entire separatrix streamline with $\psi = 0$,
i.e., $\gamma_b \cup \gamma_z = \gamma_0$, cf.~\eqref{eq:Cs}, in cost
functional \eqref{eq:J}. Since contour $\gamma_0$ is closed, by
Stokes' theorem, measurements $m$ of the tangential velocity determine
the total circulation $\Gamma$ contained in the region $\Omega$. Thus,
the reconstruction problem formulated in this way is {quite} complete.

In order to assess the effect of the initial guess $f_0$ on the
convergence of gradient algorithm \eqref{eq:desc}, we analyze
iterations staring from two distinct initial guesses $f_0$, one
underestimating and one overestimating the exact vorticity function
\eqref{eq:FHill} (since these initial guesses are representative of a
broad range of functions with similar structure, the exact formulas
are not important). The Sobolev gradients are computed using Neumann
boundary condition \eqref{eq:gradJH1ln} at the left endpoint ($s=0$)
of the identifiability interval $\I$. {The reason is that using
  instead} {homogeneous} Dirichlet boundary condition
\eqref{eq:gradJH1ld} together with $f_0(0) = C = 1/2$ would make the
reconstruction problem too easy, whereas imposing $f_0(0) \neq 1/2$
would be inconsistent with measurements $m$.  {In the present problem
  uniformly positive reconstructions $\hat{f}$ were obtained for the
  vorticity function without any positivity enforcement.}

\begin{figure}
\begin{center}
\includegraphics[width=0.8\textwidth]{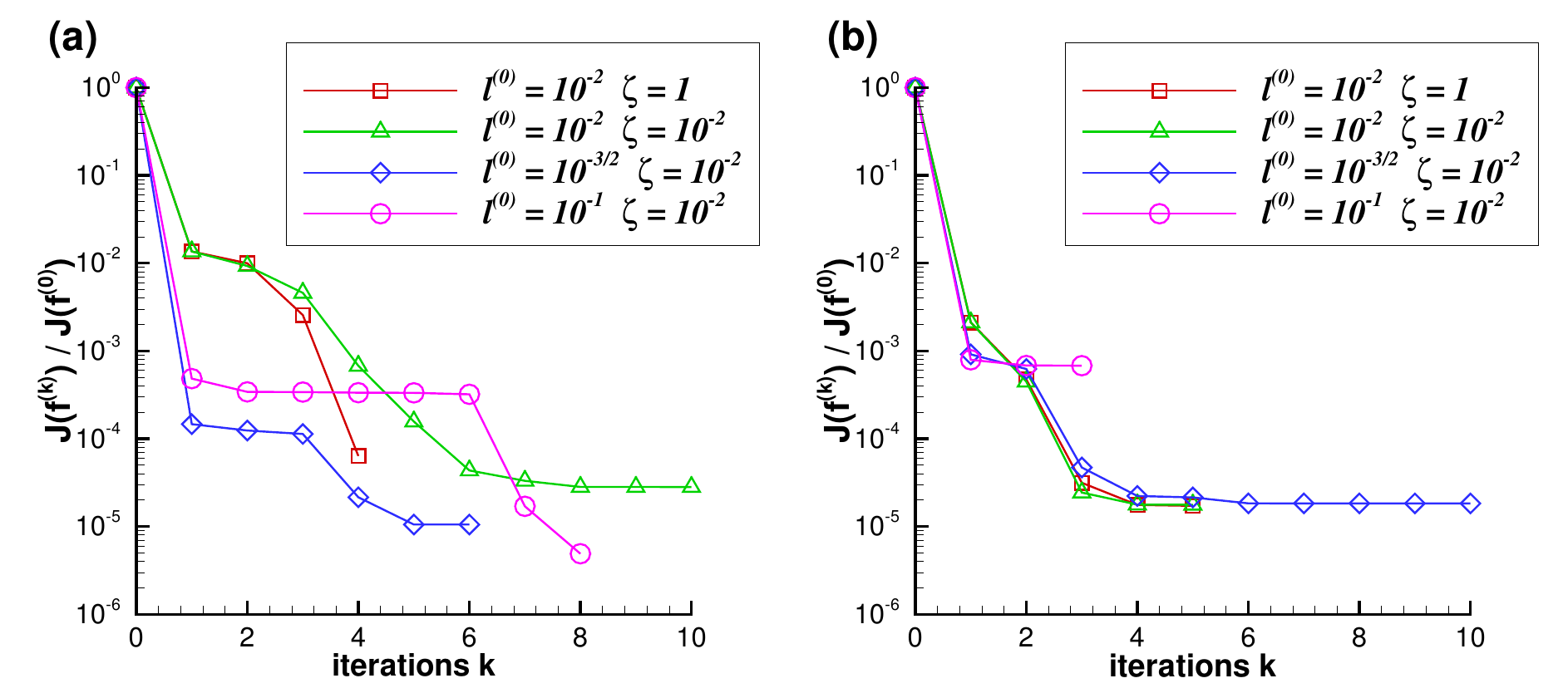}
\vspace*{-0.3cm}
\end{center}
\caption{[Hill's vortex] Decrease of cost functional $\J(f^{(k)})$
  with $k$ for iterations starting with initial guess {$f_0$ (a)
    underestimating and (b) overestimating the exact vorticity
    function \eqref{eq:FHill}}.  Different lines correspond to the
    values of $\ell^{(0)}$ and $\zeta$, cf.~equation \eqref{eq:zeta},
    indicated in the figure legends.}
\label{fig:JHill}
\end{figure}

The histories of cost functional $\J(f^{(k)})$ with iterations
corresponding to the two initial guesses are shown in Figures
\ref{fig:JHill}a and \ref{fig:JHill}b, where we consider cases with
different $\ell^{(0)}$ and $\zeta$ ({cf.~}formula
\eqref{eq:zeta}). {As discussed in Section
  \ref{sec:grad}\ref{sec:sobolev}, the values of the length-scale
  parameter $\ell^{(0)} \in [10^{-2},10^{-1}]$ are selected to lie
  within the range of variation of the streamfunction $\psi$ which in
  this problem is $[0,0.2]$.}  We note that in most cases the cost
functional decreases by about five orders of magnitude over a few
iterations.  Reducing the length-scale $\ell^{(k)}$ with iterations
has an effect on the rate of convergence when {one} initial guess is
used, but appears to play little role when the other initial guess is
employed.  Hence, in this problem, we will adopt the values
$\ell^{(0)} = {10^{-1}}$ and $\zeta = {10^{-1/5}}$. This
choice of $\zeta$ ensures that $\ell^{(k)}$ decreases by an order of
magnitude every five iterations, cf.~\eqref{eq:zeta}.

In Figures \ref{fig:fHill}a and \ref{fig:fHill}b we show the optimal
reconstructions $\hat{f}$ obtained in the two cases together with the
corresponding initial guesses.  We observe that in both cases the
reconstructed vorticity function $\hat{f}$ is very close to the exact
solution $C=1/2$ on the interval $[0, \psi_{\text{max}}]$, where
$\psi_{\text{max}} = 0.2$. {In Figure \ref{fig:fHill}b we note a
  slight deviation of the reconstructed vorticity function from the
  exact profile $f(\psi) = C = 1/2$ for values of $\psi$ close to
  $\psi_{\text{max}}$. Given that the corresponding value of cost
  functional \eqref{eq:J} is $\O(10^{-7})$, this provides evidence for
  a degree of ill-posedness of Problem \ref{P1}, in the sense that
  finite modifications of the vorticity function $f$ have only a
  vanishing effect on the measurements appearing in \eqref{eq:J}. From
  the physical point of view, this behavior can be attributed to the
  fact that streamfunction values close to $\psi_{\text{max}}$ are
  attained on a small part of the domain $\Omega$ which is close to
  the centre of the vortex and therefore removed from the contour
  $\gamma_0$ where the measurements are acquired. If the goal is to
  maximize the reconstruction accuracy for values of $\psi$ close to
  $\psi_{\text{max}}$, this issue can be remedied by including
  additional measurements acquired within the vortex bubble $\Omega_b$
  in cost functional $\J(f)$. As discussed in Section
  \ref{sec:grad}\ref{sec:min}, the reconstruction method does allow
  for such a possibility.}

The convergence of circulation $\Gamma$, impulse $I$ and energy $E$
with iterations $k$ to the values characterizing the exact solution is
shown in Figures \ref{fig:diagHill}a and \ref{fig:diagHill}b for the
two cases.  In these figures we plot the relative error
\begin{equation}
    \varepsilon^{(k)}(\Gamma):=\Bigg|\frac{\Gamma(f^{(k)})}{\Gamma_{\text{Hill}}}
    - 1\Bigg|
\label{eq:err}
\end{equation}
for the vortex circulation and analogous expressions for the impulse
and energy using logarithmic scale to determine the number of
significant digits captured in the reconstruction. In the figures we
note a fast, {though nonmonotonous, convergence} of the three
diagnostic quantities to the corresponding exact values. We obtain
approximately two digits of accuracy for the energy and three or more
for the circulation and impulse.

\begin{figure}
\begin{center}
\includegraphics[width=0.8\textwidth]{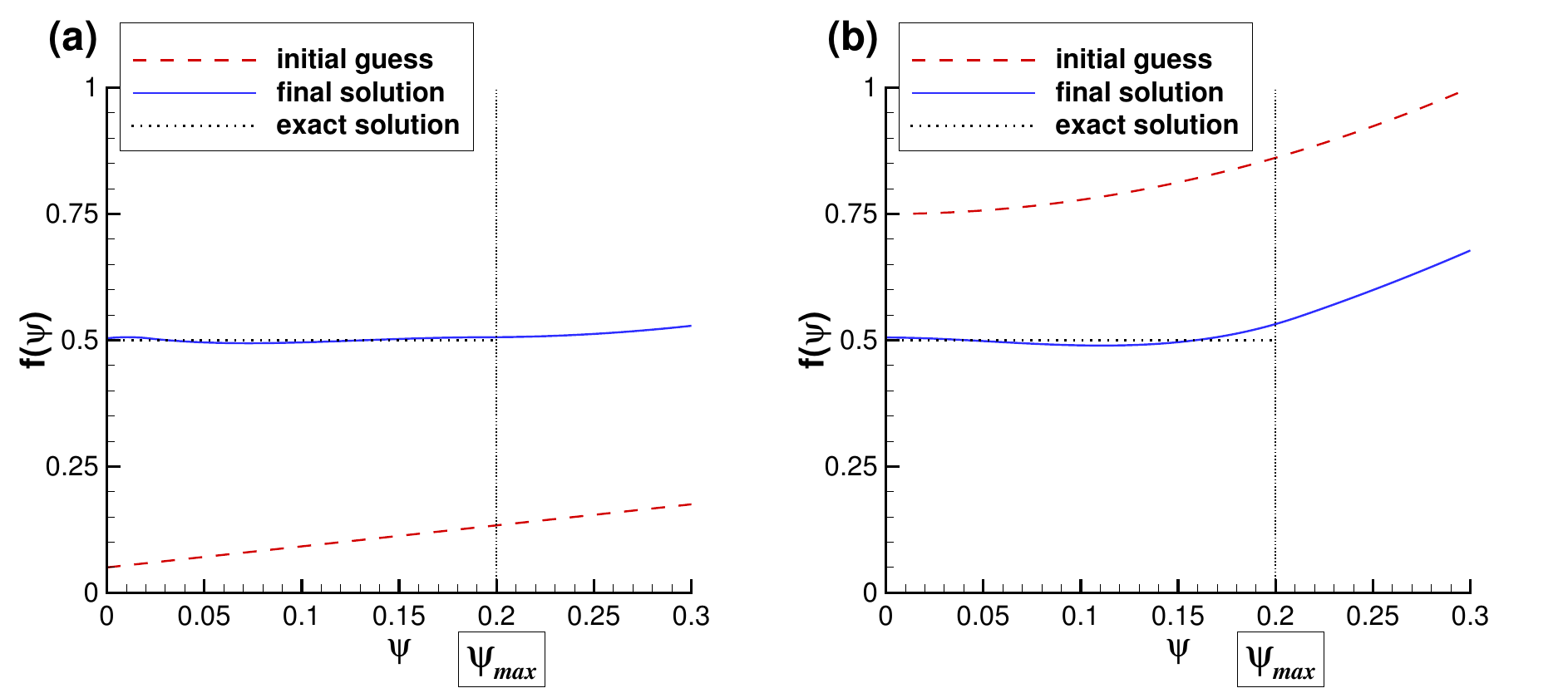}
\vspace*{-0.3cm}
\end{center}
\caption{[Hill's vortex] Reconstructed source functions $\hat{f}$
  ({blue} solid lines) and the corresponding initial guesses ({red}
  dashed lines) {when $f_0$ (a) underestimates and (b) overestimates
    the exact vorticity function \eqref{eq:FHill}}.  The black
  {horizontal dotted} line represents exact solution \eqref{eq:FHill}
  with $a=2$ and $C=1/2$, whereas the vertical dotted lines mark the
  maximum value {$\psi_{\text{max}} = 0.2$} achieved by the
  streamfunction in exact solution.}
\label{fig:fHill}
\end{figure}

\begin{figure}
\begin{center}
\includegraphics[width=0.8\textwidth]{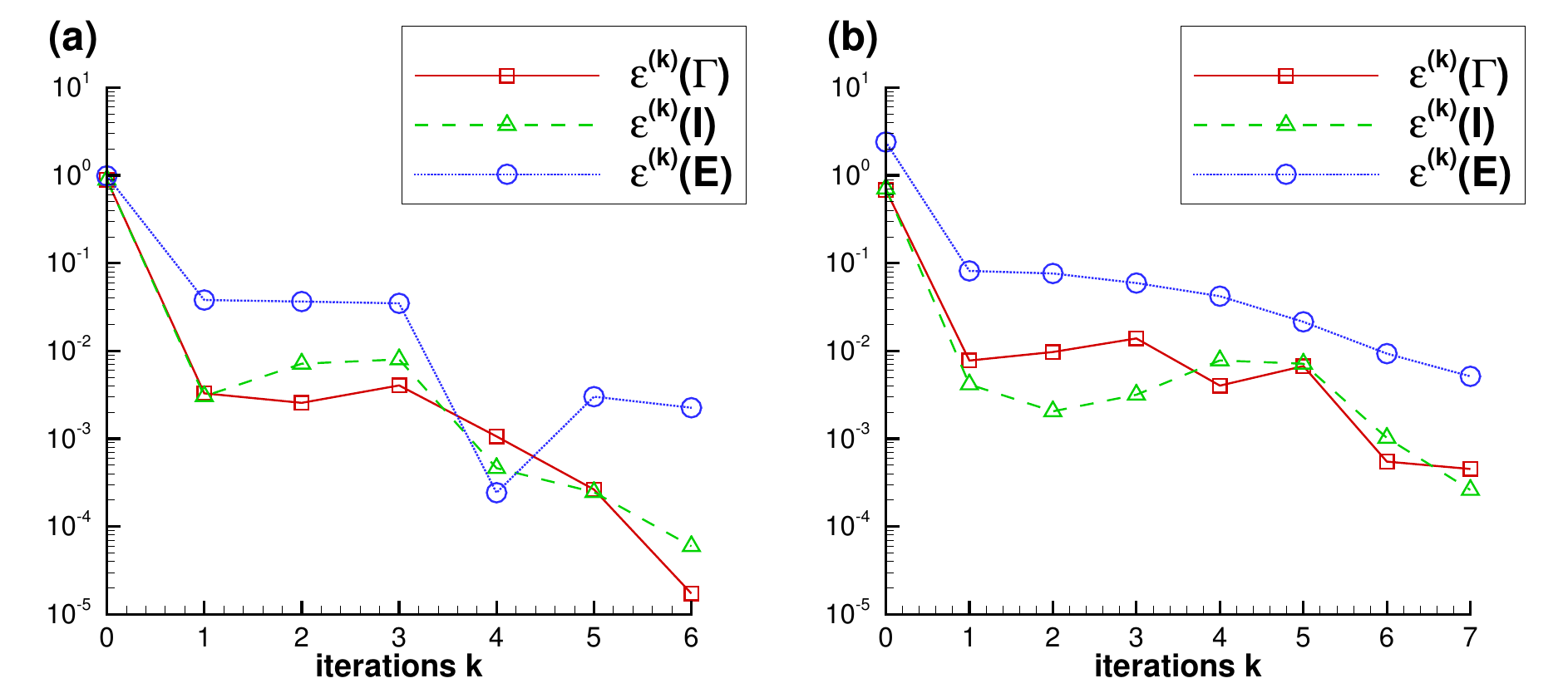}
\vspace*{-0.3cm}
\caption{[Hill's vortex] Evolution of the relative error
  $\varepsilon^{(k)}(\Gamma)$ for the circulation,
  {cf.~\eqref{eq:err},} and {of analogously defined} expressions for
  the impulse $I$ and energy $E$ showing convergence to the
  corresponding values $\Gamma_{\text{Hill}}$, $I_{\text{Hill}}$ and
  $E_{\text{Hill}}$ characterizing the exact solution,
  {cf.~\eqref{eq:GIEHill}}.  Iterations starting with {initial
    guess $f_0$ underestimating and overestimating the exact vorticity
    function \eqref{eq:FHill} are shown in panels (a) and (b),
    respectively.}}
\label{fig:diagHill}
\end{center}
\end{figure}

\section{Reconstruction of Vorticity  from DNS Data}
\label{sec:DNS}

In this section we describe {a more} challenging task of
reconstructing {the vorticity function characterizing} a realistic
vortex ring.  {In the following we use a high-resolution DNS of the}
axisymmetric Navier-Stokes equations to generate a realistic
{evolution of a} viscous vortex ring. \revt{The numerical approach is
  described in \cite{dan-2008}, although the data used in the present
  study corresponds to a higher Reynolds number and injection
  parameters chosen to reproduce the experiments reported in
  \cite{stewart-2012-ExpF}, see also \cite{dan-2009}. The
  computational domain $(z,r) \in [0,10] \times [0,2]$ is discretized
  with $3200 \times 800$ grid points ensuring convergence of the
  results.}  The vortex ring is generated by prescribing an
appropriate axial velocity profile at the inlet section of the
computational domain. We used the specified discharge velocity (SDV)
model proposed in \cite{dan-2009} to mimic the flow generated by a
piston/cylinder mechanism pushing a column of fluid through a long
pipe of diameter $D$.  In the following, all presented quantities will
be normalized using $D$ as {the} length scale and the maximum (piston)
velocity $U_0$ at the entry of the pipe {as the velocity scale}. The
corresponding reference time is thus $D/U_0$.  The main physical
parameter of the flow is the Reynolds number based on the
characteristic velocity {$Re_D ={U_0\, D}/{\nu}=17,000$,} with $\nu$
the viscosity of the fluid.  \revt{Even at this elevated Reynolds
  number the vortex ring is known to remain laminar
  \cite{akhmetov-2009}.} The injection is characterized by the stroke
length ($L_p$) of the piston. We prescribed a piston velocity program
used in {actual} experiments {with $L_p/D =1.28$
  \cite{stewart-2012-ExpF}}.

For the reconstruction problem, we consider the {vortex ring data
  obtained from the DNS at the nondimensional time} {$t=10$}.
This time instant corresponds to the post-formation phase, since the
injection stopped at $t_{off}=2.26$.  The DNS streamfunction
$\psi_{\text{DNS}}{(z,r)}$ in the frame {of reference} moving with the
vortex is computed by solving the general equation
\eqref{eq-gen-psiom} within the rectangular domain used for {the DNS
  together with the corresponding boundary conditions}.  We then use
the level set $\psi_{\text{DNS}}=0$ to define the reconstruction
domain $\Omega$ (see Figure \ref{fig-intro-dns}b) and from this data
extract the measurements {$m=m(z,r)$} on $\gamma_b$ and
$\gamma_z$, which serve as the target data in optimization problem
\eqref{eq:J}--\eqref{eq:min}. {In the following, the velocity
  field is scaled by the translation velocity $W$ and distances are
  scaled by the vortex radius. The vortex centre is thus located at
  $(z,r) = (0,1)$ and $0 \leq \psi_{\text{DNS}} \leq 0.791$.}

\subsection{Results of our reconstruction method}

As a starting point, the empirical relation $\{ \omega(z_p,r_p) / r_p,
\psi(z_p,r_p)\}_p$ between the vorticity and the streamfunction,
cf.~\eqref{eq-om-fpsi}, at the points $(z_p,r_p)$ discretizing the
flow domain $\Omega$ is shown as a scatter plot in Figures
{\ref{fig:fdns}a,b}. {While these points tend to cluster along a
  rather well-defined curve, their local scattering} is a
manifestation of the fact that the original Navier-Stokes flow is
viscous and not strictly steady in the chosen frame of reference.
\revt{This scatter tends to increase for vortex rings with smaller
  Reynolds numbers.}  {We now consider two approaches to
  reconstructing the vorticity function $f$ on the RHS in
  \eqref{eq:Euler2Da} so that the {inviscid} vortex-ring model
  provides an accurate representation of the DNS data as quantified by
  the cost functional \eqref{eq:J}.}

\begin{figure}[h]
\begin{center}
\mbox{
\subfigure[]{\includegraphics[width=0.45\textwidth]{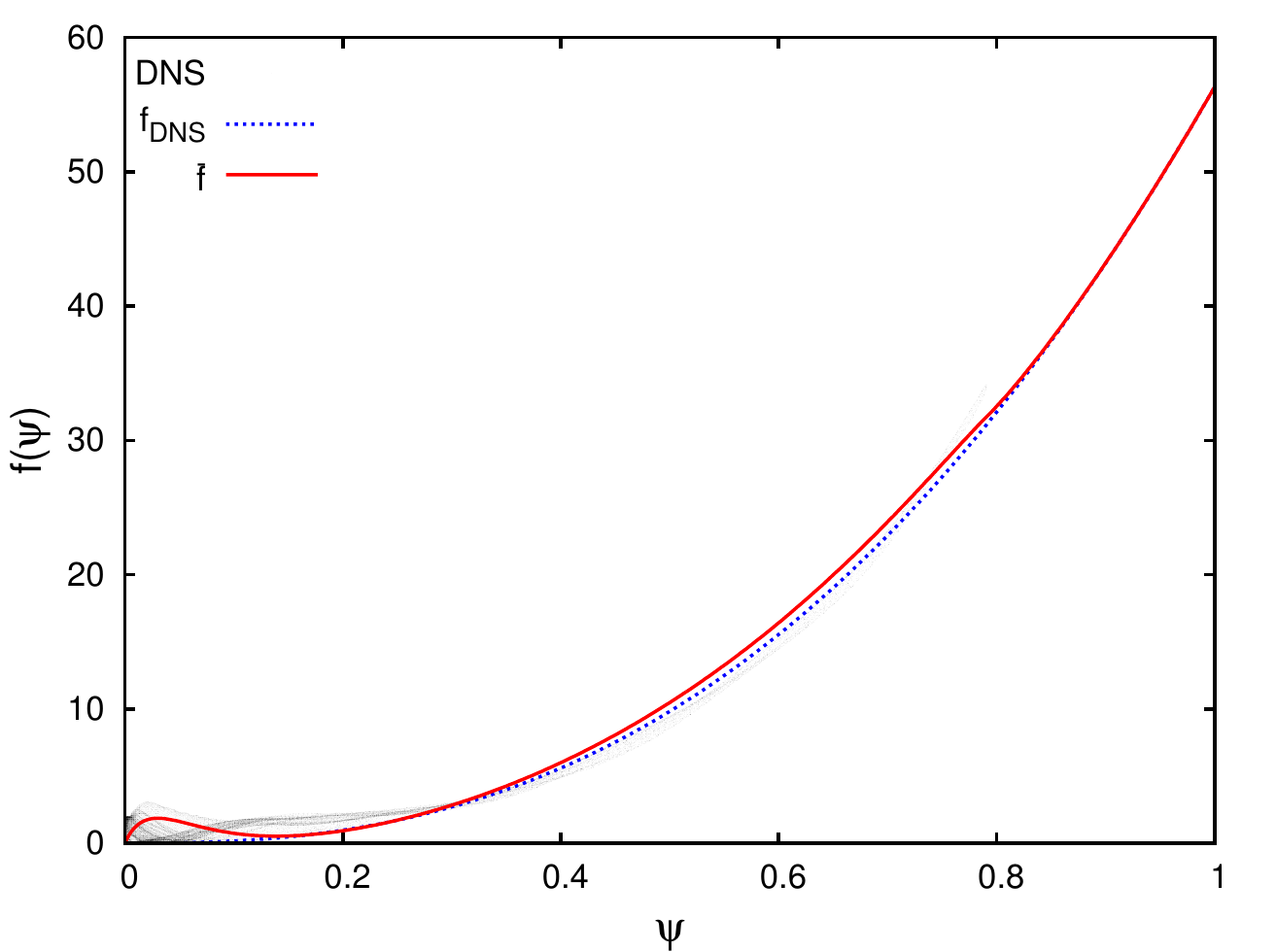}} \quad
\subfigure[]{\includegraphics[width=0.45\textwidth]{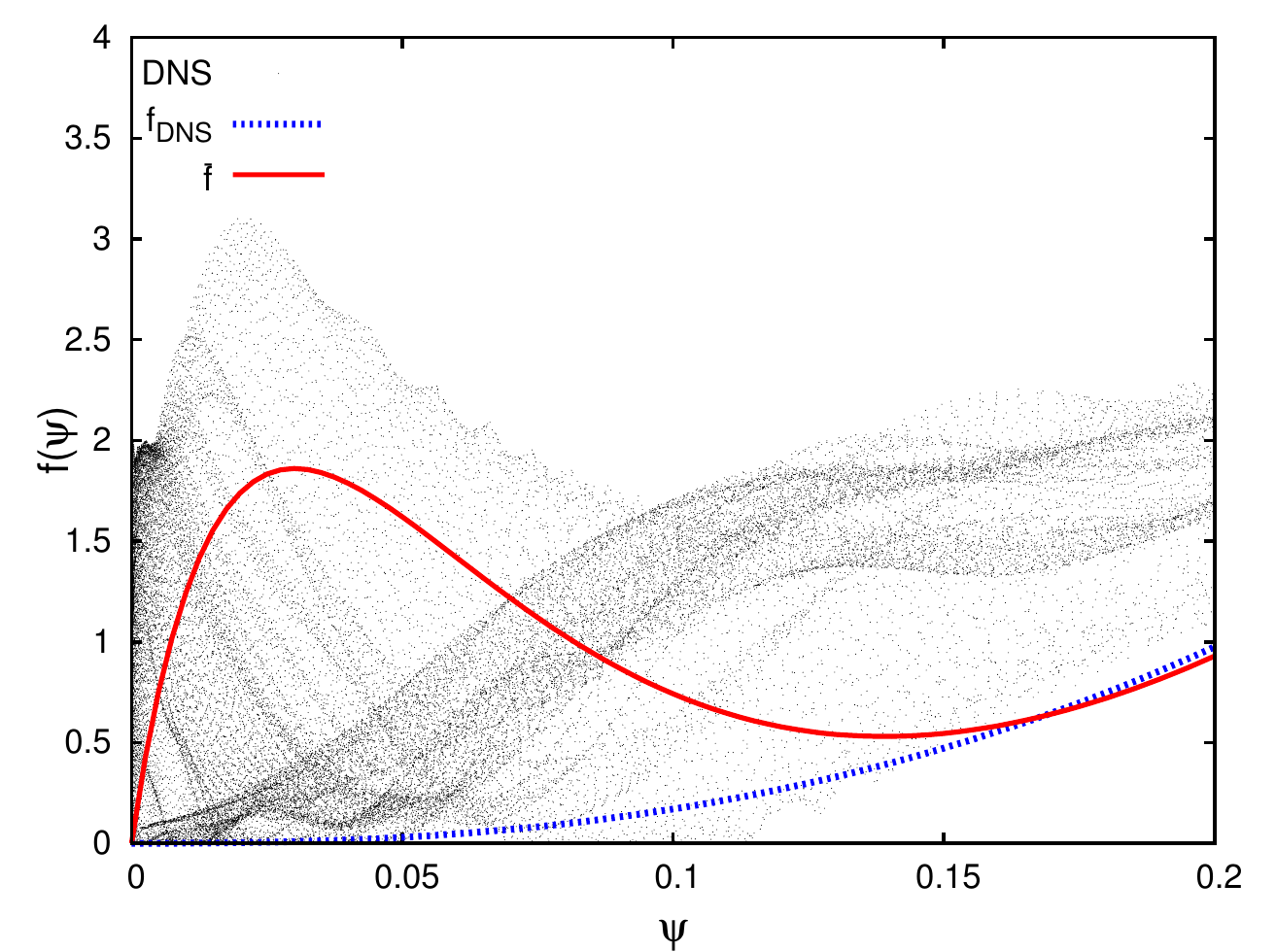}}}
\vspace*{-0.75cm}
\end{center}
\caption{[DNS] Vorticity function $f(\psi)$ {for (a) $\psi \in
    \I$ and (b) $\psi \in [0,0.2]$ (magnification of the region near
    the origin)}: (scatter plot with {dots}) relation $\{
  \omega(z_p,r_p) / r_p, \psi(z_p,r_p)\}_p$ corresponding to discrete
  Navier-Stokes DNS data; {(blue dotted line)} least-squares fit
  $f_{DNS}$ given by \eqref{eq:fdns} and (red solid line) the optimal
  reconstruction $\hat{f}$.}
\label{fig:fdns}
\end{figure}

{As the first approach we examine a least-squares fit of an
  empirical power-law relation for the vorticity function which yields
\begin{equation}
f_{\text{DNS}}(\psi) = 56.337 \, \psi^{2.520}.
\label{eq:fdns}
\end{equation}
We note that this relation has the property $f_{\text{DNS}}(0) = 0$
ensuring that the vorticity support in the inviscid vortex-ring model
coincides with $\Omega$. As is evident from Figure \ref{fig:fdns}a,
the fit captures the main trends exhibited by the data, except in the
neighbourhood of the origin where the data reveals a larger scatter.
On the other hand, due to its functional form the fit
$f_{\text{DNS}}(\psi)$ approaches zero monotonously as $\psi
\rightarrow 0$ (cf.~Figure \ref{fig:fdns}b). In addition, the
empirical fit \eqref{eq:fdns} also slightly misrepresents the slope
$df/d\psi$. From the point of view of our reconstruction problem, the
region characterized by small values of $\psi$ is particularly
important, because it involves many data points in close proximity to
the contour $\gamma_0$ on which the measurements are defined (this is
reflected by a larger density of data points near the origin in
Figures \ref{fig:fdns}a,b). The values of the cost functional and the
normalized errors in the reconstruction of the circulation, impulse
and energy, cf.~\eqref{eq:G}--\eqref{eq:E}, corresponding to the
empirical fit \eqref{eq:fdns} are collected in Table \ref{tab:dns}.

\begin{table}[h]
\begin{center}
{
\begin{tabular}{l||l|l|l|l|l|}
    & $\J(f)$ & $\Gamma$ & $I$ & $E$ & $\omega_{\max}$  \\ \hline
\Bmp{4.25cm} \smallskip DNS \smallskip \Emp  & & 4.670 & 13.393 & 7.444 & 34.07 \\ \hline
\Bmp{4.25cm} \smallskip $f_{\text{DNS}}$ \smallskip \Emp & 0.03263 & 4.212  & 13.002 & 7.470 & 32.17 \\ \hline
\Bmp{4.25cm} \smallskip $\hat{f}$  \smallskip \Emp & 0.00315 & 4.607 & 13.385  & 6.900 & 29.19  \\ \hline
\Bmp{4.25cm} \smallskip Norbury-Fraenkel model  \smallskip \Emp & 0.00864 & 4.671  & 13.974   & 7.448 & 14.45 \\ \hline
\Bmp{4.25cm} \smallskip Kaplanski-Rudi model  \smallskip \Emp &   0.07939 & 3.898  & 12.288   & 6.647 & 30.43\\ \hline
\end{tabular}
}
\end{center}
\vspace*{-0.4cm}
\caption{[DNS] Values of the cost functional \eqref{eq:J}, the 
  diagnostic quantities \eqref{eq:GIE} and the 
  {vorticity maximum $\omega_{\max}$} obtained for the 
  {empirical fit $f_{\text{DNS}}$ of the} vorticity function, 
  cf.~\eqref{eq:fdns}, the optimally reconstructed vorticity function $\hat{f}$ and the fits
  with the Norbury-Fraenkel and Kaplanski-Rudi vortex-ring models (cf.~Section \ref{sec:DNS}\ref{sec:NFKR}).}
\label{tab:dns}
\end{table}

As the second approach, we reconstruct the vorticity function
optimally by solving Problem \ref{P1} as described above, with the
least-squares fit \eqref{eq:fdns} used as the initial guess in
algorithm \eqref{eq:desc}, i.e., $f_0 = f_{\text{DNS}}$. As discussed
in Section \ref{sec:grad}\ref{sec:sobolev}, in the absence of other
relevant information, this fit will also be used to determine the
behavior of the optimal reconstruction $\hat{f}(\psi)$ for limiting
values of $\psi$. More precisely, in the reconstruction process we use
the Sobolev gradients determined subject to the homogeneous Dirichlet
boundary condition at $\psi=0$, cf.~\eqref{eq:gradJH1_2}, which
together with our choice of the initial guess ensures that $\hat{f}(0)
= 0$. As regards large values $\psi > \max_{\x \in \Omega} \psi(\x)$,
we will assume that the slope of $\hat{f}(\psi)$ will be given by the
slope of the fit \eqref{eq:fdns} which is ensured by the use of the
homogeneous Neumann boundary condition \eqref{eq:gradJH1_3} at $\psi =
\psi_{\text{max}}$. The characteristic length-scale in the inner
product \eqref{eq:ipH1} was chosen as $\ell = 10^{-3/2} \approx
0.0316$ and the decrement, cf.~\eqref{eq:zeta}, as $\zeta = 10^{1/5}$.
The optimal vorticity function $\hat{f}(\psi)$ reconstructed in this
way is shown in Figures \ref{fig:fdns}a,b, whereas the corresponding
values of the cost functional \eqref{eq:J} and the diagnostic
quantities \eqref{eq:GIE} are indicated in Table \ref{tab:dns}. First
of all, we observe that the reconstruction error as measured by the
cost functional is reduced by an order of magnitude. This is achieved
with an optimal vorticity function $\hat{f}(\psi)$ exhibiting a local
maximum around $\psi = 0.03$ which allows it to better match data
points (more scattered for small $\psi$) and at the same time satisfy
the constraint $\hat{f}(0) = 0$.  This is facilitated by our choice of
the cost functional which is more sensitive to points in $\Omega$
characterized by small values of $\psi$ as they are located close to
the boundary $\gamma_0$, cf.~Figure \ref{fig-intro-dns}. On the other
hand, for larger values ($\psi \approx 0.8$), the optimal
reconstruction reveals only a small improvement with respect to the
empirical fit $f_{\text{DNS}}$. This is a consequence of the fact that
parts of the flow domain with large values of $\psi$ are rather far
from the contour $\gamma_0$ where the measurements are acquired. Given
that the reconstruction errors represented by the cost functional
$\J(f)$ are already very small, cf.~Table \ref{tab:dns}, this effect
can be attributed to the ill-posedness of the underlying inverse
problem. As already discussed in Section \ref{sec:results}, further
improvements can be obtained using measurements distributed inside the
vortex. As documented in Table \ref{tab:dns}, the optimal
reconstruction improves the relative accuracy of both circulation and
impulse by about one order of magnitude with respect to the model
based on the empirical fit \eqref{eq:fdns}.} \revt{On the other hand,
the latter approach captures the energy and maximum vorticity (which
are both nonlinear functions of the flow variables) more accurately
than the optimal reconstruction.}

\subsection{Comparison with reconstruction methods based on analytical
  models}
\label{sec:NFKR}

In this section we compare our results with more classical
reconstruction approaches based on fits with analytical vortex-ring
models. The approach directly related to our method relies on the
Norbury-Fraenkel (NF) model for the steady inviscid vortex ring.  It
considers a constant vorticity function given by (\ref{eq-gen-frz})
and, once the vortex bubble was fixed, there are two parameters
defining the vortex ring in this model: the vorticity intensity $C$
(\ie $f(\psi)=C$) and the flux constant $k$ ($2\pi k$ represents the
flow rate between the axis $0z$ and the boundary $\partial \Omega_c$,
see Figure \ref{fig-intro-dns}b). It was shown in \cite{dan-2013-AMM}
that, for a fixed vortex bubble, vortex-ring solutions exist if $C/k
\geq \delta_{max}$, with $\delta_{max}$ estimated as a function of the
first eigenvalue of the operator $\cal L$ on the bubble domain
$\Omega$. If the vortex ring circulation is imposed, the solution is
then unique and can be numerically calculated by an iterative
algorithm suggested in \cite{dan-2013-AMM}. In our case, corresponding
to Figure \ref{fig-intro-dns}, the fitting procedure gives the following
values of the parameters: $C=13.39$ and $k=0.333$.

In the NF model, the vorticity distribution in the vortex core is
linear, \ie proportional to the distance from the symmetry axis (as in
Hill's spherical vortex model). This is not realistic, since a
Gaussian vorticity distribution was reported in experimentally
generated vortex rings \cite[\eg][]{weigand-1997-ExpF,cater-2004}.
This is remedied in the Kaplanski-Rudi (KR) vortex-ring model
\cite{kaplanski-2005-PF} which was derived as a linear first-order
solution to the Navier-Stokes equation in the axisymmetric geometry
and arbitrary times (see also \cite{kaplanski-1999-IJFMR}). The
vorticity in the vortex core was predicted to be quasi-Gaussian,
expressed by
\begin{equation}
\omega(z,r)= \frac{\Gamma_0}{\sqrt{2 \pi}} \frac{\theta^3 }{R_0^2} \exp 
\left[-\frac{1}{2}\left( \left(\frac{r}{L}\right)^2+ \left(\frac{z}{L}\right)^2 +\theta^{2}\right)\right]
\textup{I}_{1} \left( \frac{r}{L} \theta \right),
\label{eq-KR-vort}
\end{equation}
where $\textup{I}_{1}$ is the modified Bessel function of the first
kind, $L$ the effective viscous length scale of the vortex ring,
$\Gamma_0$ the circulation of the vortex ring, $R_0$ its radius ($OC$
in Figure \ref{fig-intro-dns}b) and $\theta= R_0/L$ a (viscous) parameter
identifying the vortex. To use the KR model, we fit the DNS vorticity
field with distribution \eqref{eq-KR-vort} using the approximation
$I_1(\eta) \approx \exp(\eta)/\sqrt{2\pi \eta}$ for large $\eta$.
Under this approximation relation \eqref{eq-KR-vort} becomes an
isotropic 2D Gaussian. Using a non-linear fit with the BFGS
minimization method we obtain the following values of the model
parameters: $\Gamma_0=3.925$, $R_0=1.0047$, $L=0.142$, and
$\theta=7.03$.  We note that the fitted radius of the vortex ring is
very close to the DNS value $R_0^{\text{DNS}}=1$ imposed by the
adopted scaling (see Figure \ref{fig-intro-dns}).

In our analysis we will focus on the vorticity fields of the different
models as they exhibit more significant variation than the
corresponding streamfunction fields.  Vorticity contours for the DNS
vortex ring and corresponding reconstructed fields are shown in Figure
\ref{fig:DNScompOM}. For our optimal reconstruction, at each point
$(z,r)$ the vorticity is computed as $\omega(z,r)= r
\hat{f}(\psi(z,r))$ using cubic spline interpolation, cf.~Section
\ref{sec:comput}\ref{sec:modules}. The maximum values of the
reconstructed vorticity $\omega_{\max} := \max_{(z,r) \in \Omega}
\omega(z,r)$ are given in table \ref{tab:dns}. They correspond to the
vorticity at the centre of the vortex ring, except for the NF model in
which $\omega(z,r)=C r$. One can see that our model and the viscous KR
model well approximate $\omega_{\max}$ from the DNS field.  In Figure
\ref{fig:DNScompOM} it is also interesting to note that the prolate
isocontour shapes in the DNS vorticity field are reproduced in our
model. This is neither the case for the NF model (in which the
vorticity isolines are described by $r=const$), nor for the viscous KR
model (which features quasi-circular vorticity contours, as expected
from formula \eqref{eq-KR-vort} and its approximation by an isotropic
Gaussian). \revt{As regards the KR model, we remark that this issue is
  remedied in its generalization proposed in \cite{fukumoto-2008},
  although it cannot be derived directly from the Navier-Stokes
  equation. The above} observations are also corroborated by Figure
\ref{fig:DNScompPROF} comparing the streamfunction and vorticity
profiles along different directions in the DNS data and in the
different vortex models. We note, in particular, that the vorticity
profiles corresponding to the NF model are unrealistic.

\begin{figure}[!h]
\begin{center}
\includegraphics[width=0.75\textwidth]{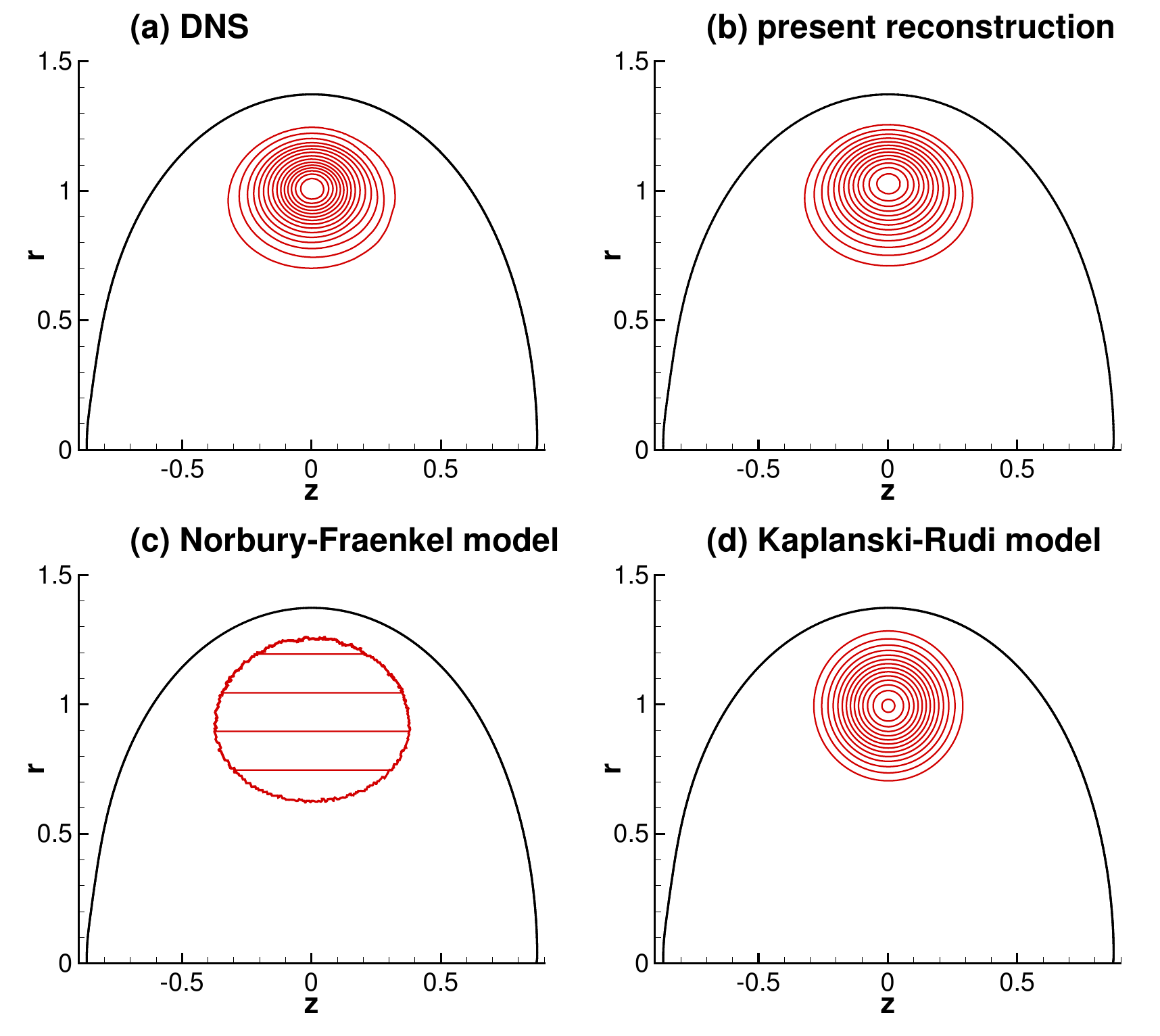}
\vspace*{-0.3cm}
\caption{[DNS] Vorticity distribution inside the vortex ring bubble
  $\Omega$: (a) DNS, (b) $\omega(z,r)= r \hat{f}(\psi(z,r))$ with the
  optimal reconstruction $\hat{f}$ of the vorticity function \revt{and
    $\psi(z,r)$ obtained as the corresponding solution to problem
    \eqref{eq:Euler2D}}, (c) fit with the Norbury-Fraenkel
    inviscid vortex-ring model, (d) fit with the Kaplanski-Rudi
    viscous vortex-ring model \eqref{eq-KR-vort}.  Vorticity
    isocontours correspond to the values $\omega = 2, 4, \dots, 32$.}
\label{fig:DNScompOM}
\end{center}
\end{figure}

\begin{figure}[!h]
\begin{center}
\includegraphics[width=0.75\textwidth]{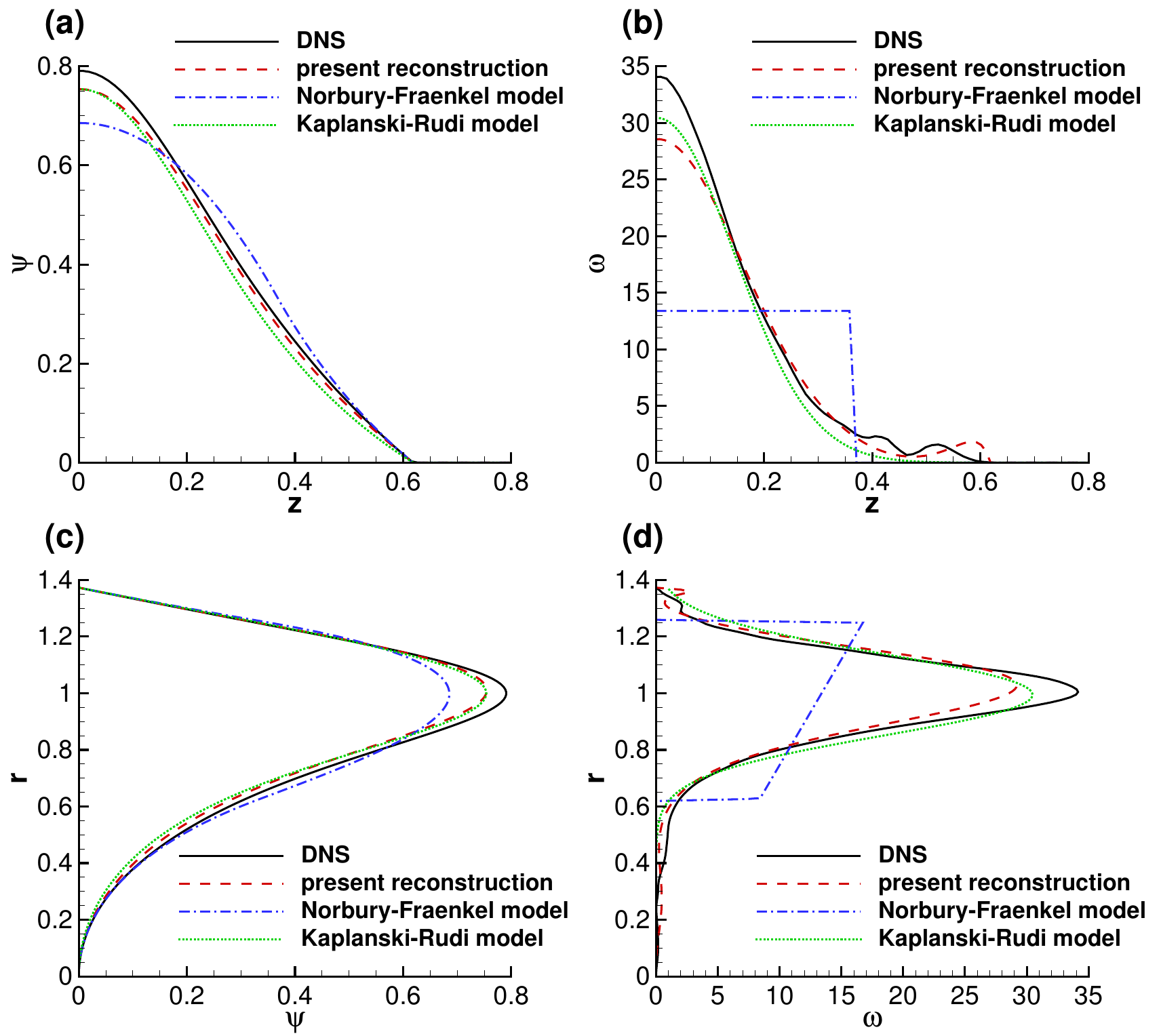}
\vspace*{-0.3cm}
\caption{[DNS] Profiles of the streamfunction $\psi$ and vorticity
  $\omega$ through the centre of the vortex ring: (a,b) along the
  $z$-axis ($r=1, z>0$; note that the profiles are symmetric with
  respect to $z=0$) and (c, d) along the $r$-axis ($z=0$; see Figure
  \ref{fig-intro-dns}b).  Comparison between the DNS data (solid
  lines), our optimal reconstruction (dashed lines) and the fits with
  the inviscid Norbury-Fraenkel model (dash-dotted lines) and with the
  viscous \revt{Kaplanski}-Rudi model (dotted lines).  }
\label{fig:DNScompPROF}
\end{center}
\end{figure}

Integral characteristics of the different vortex-ring models are
collected in Table \ref{tab:dns}.  It is interesting to note that
the NF model, which is less physical in terms of its vorticity
distribution, approximates the DNS values better than the other
models.  This explains why the NF model is often used to represent the
integral characteristics of experimentally or numerically generated
vortex rings (\eg \cite{mohseni-1998-PF,mohseni-2006}). The KR model
gives the largest errors in the integral diagnostics because the fit
used only the distribution of the vorticity, without imposing the
circulation as in the case of the NF model.

The final comparison of the different models concerns the initial
objective of the reconstruction procedure, namely, the representation
of the tangential velocity $V_t=- {(1/r)}\partial \psi / \partial
n$ on the boundary of the vortex bubble. These results are shown in
Figure \ref{fig:DNScompVT} which reveals a good agreement between the
DNS data and the predictions of the proposed and the NF model. While
the vortex model based on the optimal reconstruction of $f$ is more
accurate along the boundary $\gamma_b$ of the vortex bubble, the NF
model performs better on the axis $Oz$. In both cases the KR model
underestimates the tangential velocity which is also reflected in the
low circulation value it predicts (table \ref{tab:dns}).

\begin{figure}[!h]
\begin{center}
\includegraphics[width=0.8\textwidth]{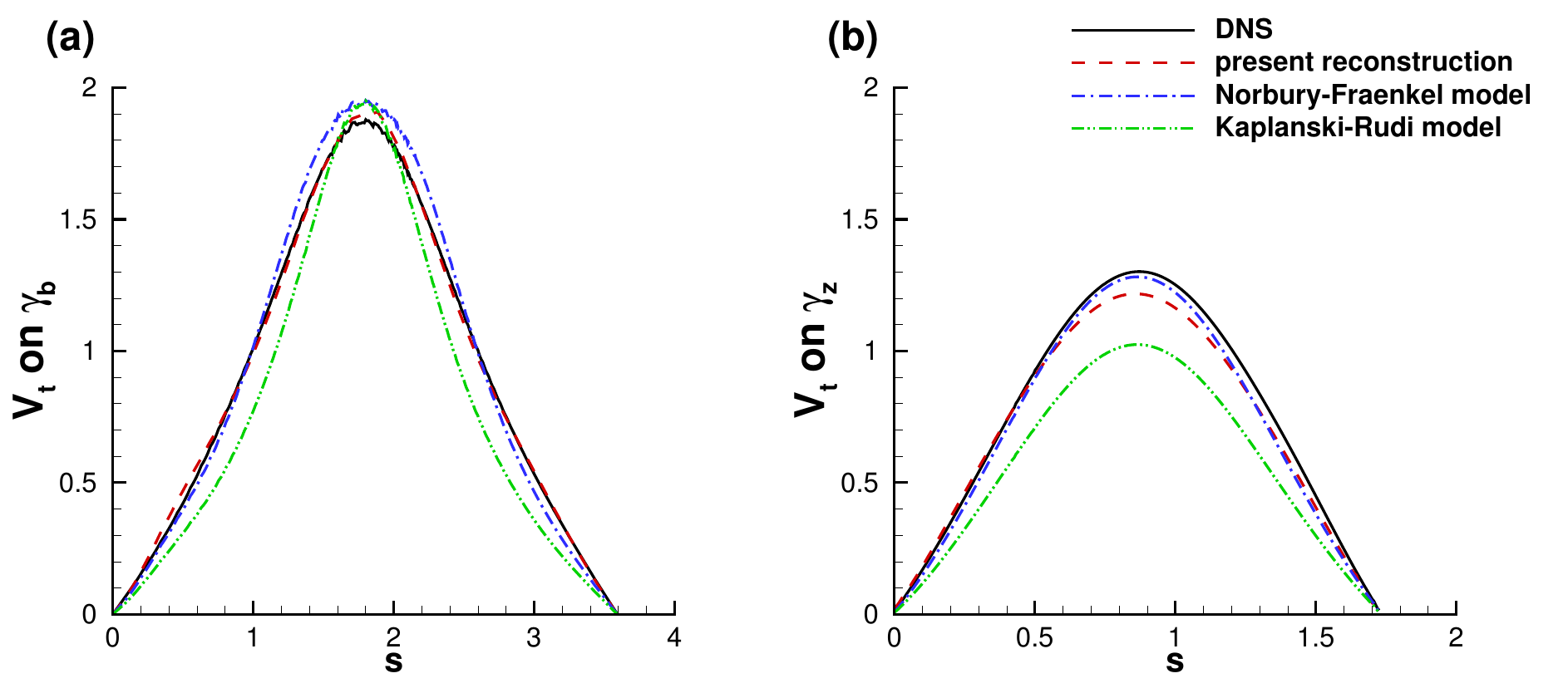}
\vspace*{-0.3cm}
\caption{[DNS] Tangential velocity on the boundary segment (a)
  $\gamma_b$ (vortex bubble) and (b) $\gamma_z$ (along the $z$-axis)
  as a function of the arc-length coordinate $s$ along the boundary
  (see also Figure \ref{fig-intro-dns}b for the definition of
  $\gamma_b$ and $\gamma_z$). Comparison between DNS data (solid
  lines), our optimal reconstruction (dashed lines), the fit with the
  inviscid Norbury-Fraenkel vortex ring model (dash-dot lines) and the
  fit with the Kaplanski-Rudi viscous vortex ring model (dotted
  lines). }
\label{fig:DNScompVT}
\end{center}
\end{figure}

We conclude that, while one of the analytic models may perform better
with respect to a particular criterion (especially the ones used to
determine its parameters), the proposed approach provides the most
balanced and consistent representation with respect to {\emph all}
considered criteria. It must also be emphasized that our optimal
reconstruction method uses information on the boundary ${\gamma_b \cup
  \gamma_z}$ of the vortex ring (see Figure \ref{fig-intro-dns}b) only,
while the reconstruction methods based on fits with the analytic
vortex models used information in the entire domain $\Omega$.

\section{Discussion, Conclusions and Outlook}
\label{sec:final}

In this study we have formulated and validated a novel solution
approach to an inverse problem in vortex dynamics concerning the
reconstruction of the vorticity function in 3D axisymmetric Euler
flows. {Solutions of such problems allow us to construct optimal
  inviscid vortex models for realistic flows.} More generally, this is
an example of the reconstruction of {a} nonlinear source term in an
elliptic PDE and as such has many applications in fluid mechanics and
beyond (more on this below). It also has some similarities to the
``Calderon problem'' which is one of the classical inverse problems
{studied in the context of} elliptic PDEs. In particular, many
questions concerning the uniqueness of the reconstructions remain
open.  In contrast to a number of earlier approaches which relied on
finite-, and usually low-dimensional, parameterizations of the
reconstructed vorticity function (e.g., \cite{dan-2012-JNM}), the
method proposed here is {\em non-parametric} and allows us to
reconstruct the vorticity function in a very general form in which
only the smoothness and boundary behavior are prescribed. A key
element of the computational approach is a suitable reformulation of
the adjoint-based optimization which was developed for the
reconstruction of constitutive relations in \cite{bvp10,bp11a} and was
already applied to other estimation problems in fluid mechanics in
\cite{pnm14,pno14a}.

In addition to standard tests on the accuracy of the adjoint gradients
presented in Section \ref{sec:comput}\ref{sec:kappa}, our approach was
validated by reconstructing the vorticity functions in {a
  classical problem involving Hill's vortex. For benchmarking
  purposes, the inverse problem was set up using a smaller amount of
  measurement data than typically available in practice making the
  reconstruction more challenging. However, its accuracy was very good
  in terms of the terminal value of the cost functional \eqref{eq:J},
  the optimal vorticity function $\hat{f}$ and the diagnostic
  quantities \eqref{eq:GIE}.  The results obtained for the case with
  the actual DNS data in Section \ref{sec:DNS} demonstrated that the
  proposed approach can significantly improve the accuracy of the
  inviscid model as compared to a simple empirical fit. This is
  achieved by obtaining a more precise representation of the vorticity
  function for small values of $\psi$, made possible by the
  nonparametric form of the reconstruction approach.  

  A thorough comparison was also made between the vortex models
  developed here and the classical models of Norbury-Fraenkel and
  Kaplanski-Rudi. \revt{Although it is more costly from the
    computational point of view,} the approach based on the optimal
  reconstruction of the vorticity function was shown to be superior in
  the sense that in addition to offering an accurate representation of
  the vorticity field inside the core and of the velocity on the
  boundary of the vortex bubble, it also captured the integral
  diagnostics with good accuracy. None of the analytic models was able
  to simultaneously achieve all of these objectives.  It ought to be
  emphasized that our approach also required significantly less
  measurement data than the NF and KR models. This good performance is
  a result of an optimization formulation central to the proposed
  approach.

  With encouraging results of the benchmark tests presented here, a
  natural research direction is to apply the proposed approach to DNS
  data and datasets obtained experimentally with techniques such as
  PIV over a broad range of Reynolds numbers. Measurement data
  available over larger parts of the flow domain should improve the
  robustness of reconstructions. \revt{The optimal reconstruction
    approach developed here will allow us to address the basic
    question how accurately actual viscous flows can be represented in
    terms of inviscid models of the type \eqref{eq:Euler2D}. In
    particular, one is interested in the fundamental limitations on
    the accuracy of such representations in terms of flow unsteadiness
    and finite viscosity effects.} An aspect of the
  \revt{reconstruction} problem which has not been addressed here, but
  which is likely to arise when using experimental data, is dealing
  with noisy measurements. This problem is however relatively well
  understood in the context of inverse problems \cite{t05}.  }

 







\section{Acknowledgements}

The authors thank two referees for their helpful comments. They are
also grateful to Dr.~Vladislav Bukshtynov for sharing some of his
software. B.~P.~acknowledges the generous hospitality of Laboratoire
de math\'ematiques Rapha\"el Salem where this work was initiated in
June 2013. He was also partially supported with an NSERC (Canada)
Discovery Grant.

 

\end{document}